\documentclass[fleqn,usenatbib]{mnras}

\usepackage{newtxtext,newtxmath}
\usepackage{caption, subcaption}
\usepackage[T1]{fontenc}

\DeclareRobustCommand{\VAN}[3]{#2}
\let\VANthebibliography\thebibliography
\def\thebibliography{\DeclareRobustCommand{\VAN}[3]{##3}\VANthebibliography}

\usepackage{graphicx}	
\usepackage{amsmath}	
\usepackage{xcolor}
\usepackage{orcidlink}

\newcommand{\Lya}[0]{Ly$\alpha$ }

\newcommand{\sm}[1]{_{\text{#1}}}
\newcommand{\Ghomog}[0]{{\bf{G}}$\sm{homog}$}

\newcommand{\Grecov}[0]{{\bf{G}}$\sm{patchy}$}

\newcommand{\edit}[1]{{{#1}}}

\newcommand{\BoeraRange}{$-2.2 \leq \text{log}_{10}[(k /(\rm{km}^{-1}~\rm{s})]\leq -0.7$} 

\newcommand{\Amatch}[0]{$A^{\text{true}}\sm{p}$}

\newcommand{\Ap}[0]{$A\sm{p}$}

          \newcommand{\skm}{\mathrm{km^{-1}\,s}} 
          
     \newcommand{\logtenk}{$\text{log}_{10}[(k /(\rm{km}^{-1}~\rm{s})]$}  
 
 \newcommand{\HI}{\hbox{H$\,\rm \scriptstyle I\ $}}
  
 \newcommand{\HeII}{\hbox{He$\,\rm \scriptstyle II\ $}}
 
\title[Large-scale enhancement in the power spectrum]{Possible evidence for a large-scale enhancement in the Lyman-$\alpha$ forest power spectrum at redshift $\mathbf{\textit{z}\geq 4}$}

\author[M. Molaro et al.] {Margherita Molaro$^{1}$\,\orcidlink{0000-0002-1841-4274}\thanks{E-mail:
    margherita.molaro@nottingham.ac.uk}, 
    Vid Ir\v{s}i\v{c}$^{2}$\,\orcidlink{0000-0002-5445-461X},
    James S. Bolton$^{1}$\,\orcidlink{0000-0003-2764-8248}\thanks{E-mail:
    james.bolton@nottingham.ac.uk}, 
    Maggie Lieu$^{1}$\,\orcidlink{0000-0002-4487-8136}, 
    Laura C. Keating$^{3}$\,\orcidlink{0000-0001-5211-1958}, 
   \newauthor Ewald Puchwein$^{4}$\,\orcidlink{0000-0001-8778-7587}, 
     Martin G. Haehnelt$^{2}$\,\orcidlink{0000-0001-8443-2393} 
    \& Matteo Viel$^{5,6,7,8}$\,\orcidlink{0000-0002-2642-5707} 
 \\$^1$School of Physics and Astronomy, University of Nottingham, University Park, Nottingham, NG7 2RD, UK
 \\$^{2}$Kavli Institute for Cosmology and Institute of Astronomy, Madingley Road, Cambridge, CB3 0HA, UK
 \\$^{3}$Institute for Astronomy, University of Edinburgh, Blackford Hill, Edinburgh,
EH9 3HJ, UK
\\$^{4}$Leibniz-Institut f\"ur Astrophysik Potsdam, An der Sternwarte 16, 14482 Potsdam, Germany\\
 $^{5}$SISSA - International School for Advanced Studies, Via Bonomea 265, I-34136 Trieste, Italy
  \\$^{6}$IFPU, Institute for Fundamental Physics of the Universe, Via Beirut 2, I-34151 Trieste, Italy
  \\$^{7}$INAF - Osservatorio Astronomico di Trieste, Via G.B. Tiepolo 11, I-34131 Trieste, Italy 
  \\$^{8}$INFN - National Institute for Nuclear Physics, Via Valerio 2, I-34127 Trieste, Italy}

\date{Accepted XXX. Received YYY; in original form ZZZ}

\pubyear{2022}

\begin{document}
\label{firstpage}
\pagerange{\pageref{firstpage}--\pageref{lastpage}}
\maketitle

\begin{abstract}
Inhomogeneous reionization enhances the 1D \Lya forest  power spectrum  on large scales at redshifts $z\geq4$.  This is due to coherent fluctuations in the ionized hydrogen fraction that arise from large-scale variations in the post-reionization gas temperature, which fade as the gas cools.  It is therefore possible to use these relic fluctuations to constrain inhomogeneous reionization with the power spectrum at wavenumbers $\log_{10}(k/{\rm km^{-1}\,s})\lesssim -1.5$.  We use the Sherwood-Relics suite of hybrid radiation hydrodynamical simulations to perform a first analysis of new \Lya forest power spectrum measurements at $4.0\leq z \leq 4.6$.  These data extend to wavenumbers $\log_{10}(k/{\rm km^{-1}\,s})\simeq -3$, with a relative uncertainty of $10$--$20$ per cent in each wavenumber bin.  Our analysis returns a $2.7\sigma$ preference for an enhancement in the \Lya forest power spectrum at large scales, in excess of that expected for a spatially uniform ultraviolet background.  This large-scale enhancement could be a signature of inhomogeneous reionization, although the statistical precision of these data is not yet sufficient for obtaining a robust detection of the relic post-reionization fluctuations.   We show that future power spectrum measurements with relative uncertainties of $\lesssim 2.5$ per cent should provide unambiguous evidence for an enhancement in the power spectrum on large scales.
\end{abstract}



\section{Introduction}

Intergalactic neutral hydrogen along the line of sight toward high-redshift quasars leaves an observable spectral signature in the form of \Lya absorption lines.  These features are cumulatively referred to as the \Lya forest \citep[see e.g.][]{Rauch1998,McQuinn2016}, and their observable properties are linked to the physical conditions in the distribution of intergalactic matter on scales of $\sim 0.5$--$50$ comoving Mpc.  The factors that determine the observable properties of the \Lya forest absorbers  -- and hence the physical properties of the intergalactic medium (IGM) -- may be grouped into two broad categories: those of cosmological origin, such as the nature of dark matter and the shape of the matter power spectrum \citep{Croft2002, Seljak2006, Irsic2017, Garzilli2019, Rogers2021, Villasenor2022}, and those of astrophysical origin, such as feedback processes \citep{Theuns2002_winds,Viel2013_feedback,Gurvich2017,Chabanier2019} and the ionization and thermal state of the IGM \citep{Schaye2000,Bolton2005,Boera2014,Hiss2018, Gaikwad2021}.  Canonically, the ionization state of the IGM is determined by the UV photons emitted by stars and active galactic nuclei. As a result, intergalactic \Lya absorption is also a key probe of the final stages of the reionization era at $z>5$ \citep{Fan2006,Becker2015,Eilers2018,Bosman2022}

A widely used statistic for characterising the \Lya forest is the 1D power spectrum of the transmitted flux \citep{McDonald2000,Palanque2013,Irsic2017_XQ100,Walther2018,Boera2019,Karacayli2022}.  When studying the IGM approaching the reionization era, the 1D power spectrum is useful in several different ways.  First, the power spectrum amplitude is sensitive to the average \Lya forest transmission, and hence also the IGM neutral hydrogen fraction \citep{Mishra2022}.  Second, the shape of the power spectrum on small scales (i.e. for wavenumbers $\log_{10} (k/\rm km^{-1}\,s) \gtrsim -1.5$) depends on the IGM thermal history, through the shape of the Doppler broadening kernel and role that gas pressure plays in setting the physical extent of \Lya absorbers \citep{Nasir2016}.  Finally, on large scales ($\log_{10} (k/\rm km^{-1}\,s) \lesssim -1.5$), the power spectrum is sensitive to spatial fluctuations in the thermal and ionization state of the IGM \citep{Cen2009}.  These fluctuations -- associated with the inhomogeneous heating of the IGM during reionization --  will linger for some time after reionization has completed due to the long cooling timescale in the low density IGM \citep{Theuns2002,HuiHaiman2003}.  The consequence is that observable relics of the reionization era will be imprinted in the \Lya forest power spectrum at redshift $z>4$.

In this context, \cite{Molaro2022} (hereafter M22) used Sherwood-Relics -- a set of hybrid radiation hydrodynamical simuluations of the IGM during reionization \citep{Puchwein2022}  -- to study the effect of patchy reionization on the 1D \Lya forest power spectrum.  In agreement with a number of other studies \citep{Cen2009,Keating2018,DAloisio2018fluc,Onorbe2019,Wu2019,Montero2020}, M22 found that remnant patches of hot, highly ionised, low density hydrogen left over following reionization produce large-scale variations in the \Lya forest transmission at $4.2<z<5$.  These enhance the 1D power spectrum of the transmitted flux by 10-50 per cent on large scales, $\log_{10} (k/\rm km^{-1}\,s) \lesssim -1.5$, with this effect being largest close to the end point of reionization.  However, M22 also demonstrated that these fluctuations will have a limited effect on the recovery of thermal parameters from \emph{existing} measurements of the \Lya forest power spectrum at $z>4$ \citep[e.g.][]{Boera2019}.  One reason for this is that the measurements of \citet{Boera2019} do not include the larger scales at $\log_{10} (k/\rm km^{-1}\,s) \lesssim -2.2$ where the additional power is expected to be most significant (see e.g. Fig. 6 in M22). Surveys such as the Dark Energy Spectropscopic Instrument (DESI) survey \citep{VargasMagana2019} and the William Herschel Telescope Enhanced Area Velocity Explorer-Quasi-stellar Object (WEAVE-QSO) survey \citep{Pieri2016} will, however, extend to such large scales with  lower resolution spectra, and will measure the 1D \Lya forest power spectrum to a precision of a few per cent.  Inhomogeneous reionization effects at $z>4$ will therefore be an important astrophysical ``nuisance'' parameter that must be included within the forward modelling frameworks used to constrain the underlying matter power spectrum from these observations.  Conversely, the presence of additional large scale power in the \Lya forest data can also provide valuable information on the end stages of inhomogeneous reionization, as this may be consistent with relic fluctuations in the ionization and thermal state of the IGM \citep[e.g.][]{DAloisio2018fluc}.

In this work we will build on M22 by investigating the possibility of detecting inhomogeneous reionization using the 1D \Lya forest power spectrum on large scales.   For this purpose we will make use of the recent 1D power spectrum measurements presented by \cite{Karacayli2022} using data from the Keck Observatory Database of Ionized Absorption toward Quasars (KODIAQ) \citep{OMeara2017}, the Spectral Quasar Absorption Database (SQUAD) \citep{Murphy2019} and
the XQ-100 survey \citep{Lopez2016}.  Importantly, these data extend to  larger scales, $\log_{10} (k/\rm km^{-1}\,s) \simeq -3$, compared to \citet{Boera2019}, although the precision is still rather modest, with relative uncertainties in the power spectrum of $10$--$20$ per cent.    As we shall demonstrate, however, this is nevertheless sufficient for providing evidence for enhanced large scale power in the data.   A related earlier result was presented by \citet{DAloisio2018fluc}, who performed an analysis of the 1D power spectrum at slightly higher redshift, $z=5.2$--$5.6$, using the sample of $21$ quasars presented by \citet{McGreer2015}.   By comparing these data with a semi-numerical model for the patchy ionization of the IGM, \citet{DAloisio2018fluc} found tentative evidence ($\sim 2\sigma$) for an enhancement in large scale power at $\log_{10}(k/{\rm km^{-1}\,s})\simeq -3$.

We furthermore introduce two important updates to the Bayesian inference framework used in M22.  First, we improve the accuracy and speed of our 1D \Lya forest power spectrum emulator, by replacing the linear interpolation approach we had previously used to build our model grid \citep[e.g.][]{VielHaehnelt2006} with a neural network.  Second, we introduce a more flexible, model independent approach for including the patchy ionization correction quantified by M22 within our Bayesian parameter estimation framework.

This paper is structured as follows.  In Section \ref{section:emulator} we briefly re-describe the simulations used in M22 and introduce the new machine-learning generated power spectrum emulator that will be adopted in this work. In Section \ref{section:Az_freeparam} we consider the effect of patchy reionization on the \Lya power spectrum, and introduce a new parameterisation that allows the power spectrum template presented by M22 to be implemented within our Markov Chain Monte Carlo analysis in a more general, model independent way.  In Section \ref{section:karac_data}, we apply our framework to the data presented by \cite{Karacayli2022} and discuss the possible evidence for an enhancement in power on large scales. We furthermore consider the possibility of applying our analysis to future, higher precision data before concluding in Section \ref{section:conclusions}.  We assume a flat $\Lambda$CDM cosmology throughout this work, with $\Omega_{\Lambda}=0.692$, $\Omega_{\rm m}=0.308$, $\Omega_{\rm b}=0.0482$, $\sigma_8=0.829$, $n_{\rm s} =0.961$, $h=0.678$ \citep{planck2014}, and a primordial helium fraction by mass of $Y_{\rm p}=0.24$ \citep{Hsyu2020}.


\section{Emulating the Lyman-alpha forest 1D power spectrum  using neural networks} \label{section:emulator}

\subsection{Numerical simulations} \label{section:grids}

In this work  -- with the exception of the neural network described in Section \ref{section:NN} -- we use the same Bayesian inference and simulation set-up adopted in M22.  This is based on a  Monte Carlo Markov Chain (MCMC) sampler combined with a Metropolis Hastings algorithm, as introduced by \cite{Irsic2017}.  The backbone of our \Lya forest power spectrum emulator consists of 12 hydrodynamical simulations that form part of the Sherwood-Relics simulation suite \citep{Puchwein2022}.  Each model follows a $20h^{-1}\rm\,cMpc$ cosmological volume with $2\times 1024^{3}$ particles using variations of the \citet{Puchwein2019} spatially homogeneous UV background synthesis model (see Table 1 in M22).   For completeness we briefly recapitulate some key information about the simulations here, but refer the reader to M22 for further details.

\begin{figure*}
    \centering
        \includegraphics[width=18cm, trim = 10 50 10 50]{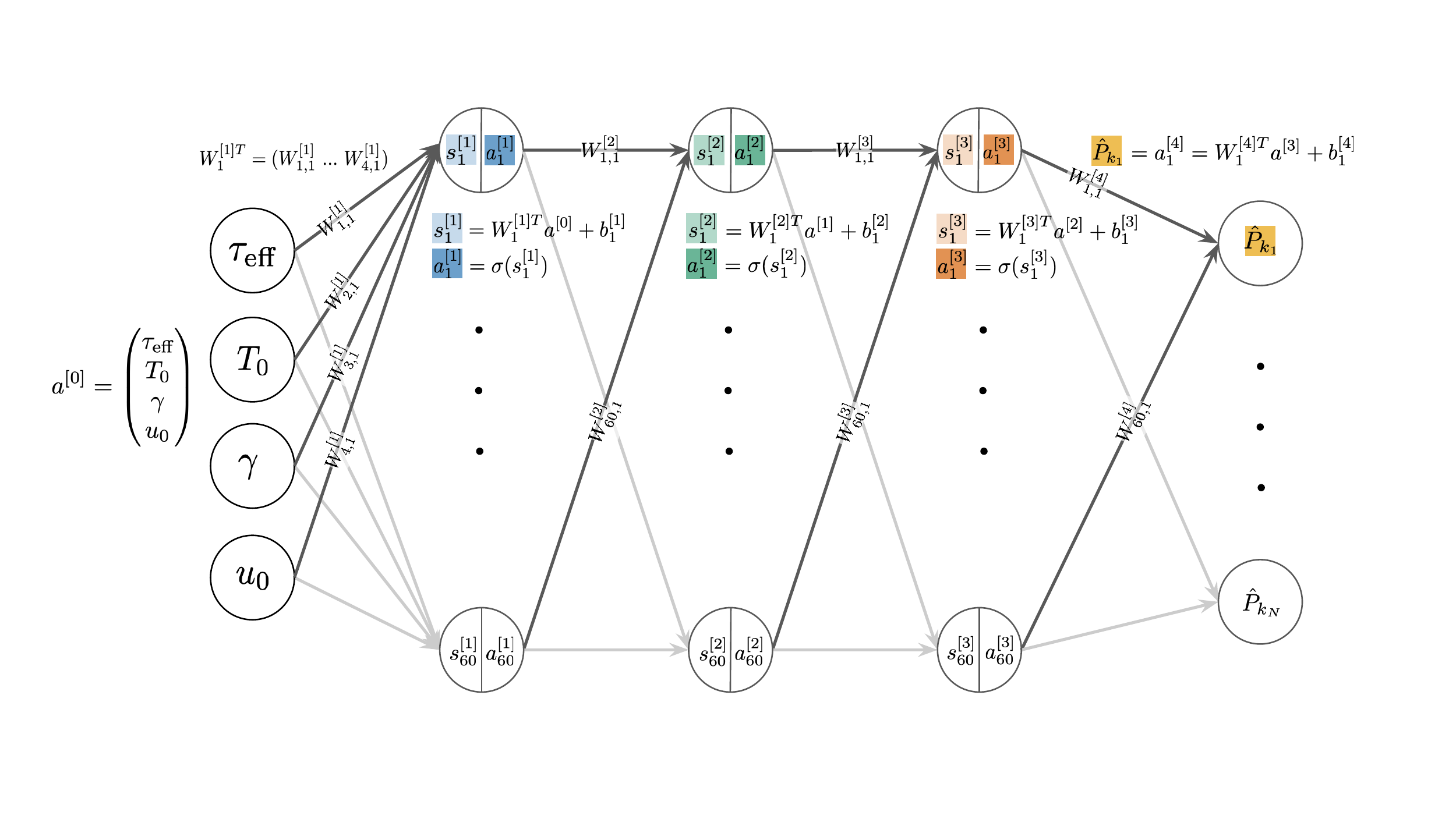} 
    \vspace{-0.7cm}
    \caption{Schematic illustrating the neural network used to predict a \Lya forest power spectrum ($\hat{P}_{k}$) sampled with $N$ bins in wavenumber, $k$, given four input astrophysical parameters ($\tau\sm{eff},T_0,\gamma, u_0$).  In addition to the input layer with four nodes for the input parameters, the network has three hidden layers (blue, green and orange) with $60$ nodes each, and an output layer (yellow) with $N$ nodes, one for each $k$-bin. $\textbf{W}$ and $\textbf{b}$ are the weight and bias matrices respectively, $\sigma$ is the ReLU activation function (see text for details), \textbf{\textit{a}} is the value of the node, and \textbf{\textit{s}} is an intermediate value used in the calculation of \textbf{\textit{a}}. The upper index in the square bracket refers to the layer (0 to 4, where 0 is the input layer and 4 the output layer), while the lower indices refer to the node. In the case of the weights, which ``connect'' two layers, the first lower index refers to the node in the first layer, and the second to the second layer.  Equations are provided only for the top node of each layer.}
    \label{fig:NN_diagram}
\end{figure*}

The IGM thermal history in each of the 12 simulations is characterised by three parameters which, as in M22, will be used to described the IGM thermal state: the cumulative energy per unit mass deposited into gas at the mean background density, $u_0$, the temperature at mean density, $T_0$, and $\gamma$, the index in the power-law relation describing the temperature-density relation, where $T = T_0 [\rho/\langle \rho \rangle]^{\gamma-1}$ \citep[][]{Hui1997,McQuinn2016}, where $\rho$/$\langle \rho \rangle$ is the local gas overdensity. The 12 simulations, each characterised by a different $u_0$ value, are then post-processed to finely sample the $T\sm{0}$ - $\gamma$ parameter-space: gas particles in the simulations are translated and rotated in the temperature-density plane to obtain different combinations of these parameters (see \cite{Boera2019}, \cite{Gaikwad2020}, and M22 for more details).

The resulting \Lya optical depths are also rescaled in post processing to give different values for the effective optical depth, $\tau\sm{eff}$, where $\tau\sm{eff} = - \text{ln} \langle F \rangle$ and $\langle F \rangle$ is the mean \Lya forest transmission.  In each redshift bin, the parameter grid uniformly spans $T\sm{0}$ between $5\,000\rm\,K$ and $14\,000\rm\,K$ in steps of $1\,000\rm\,K$, $\gamma$ between $0.9$ to $1.8$ in steps of $0.1$, and $\tau\sm{eff}$ from $0.3$ to $1.7$ times the observed effective optical depth $\tau\sm{eff,obs}$ \citep{Viel2013} in steps of $0.1\tau\sm{eff,obs}$ for each redshift bin considered.  This leads to $10 \times 10 \times 15 \times 12 = 18\,000$ post-processed sets of \Lya forest spectra in the parameter grid for each redshift considered.   The 1D \Lya forest power spectrum is then calculated from each set of spectra assuming a constant $\log_{10}(k/{\rm km^{-1}\,s})$ step-size of $0.1$, following \citet{Boera2019} and M22 (although we will adopt a different binning strategy when analysing observational data in Section~\ref{section:karac_data}).  We refer to this grid of simulated power spectra as \Ghomog{}, because it relies on simulations performed with a spatially homogeneous ultraviolet (UV) background.

Finally, as discussed in M22, for each redshift considered the $18\,000$ realisations of the power spectrum in \Ghomog{} can be modified using a template that includes the effect of inhomogeneous reionization on the large scale power at $\log_{10}(k/{\rm km^{-1}\,s})\leq -1.5$.   This template uses the data compiled in Table 2 and eq. (3) of M22 (see also Section~\ref{sec:general} later).   We refer to the resulting grid of modified power spectra as \Grecov{}, because these now also include the expected (model-dependent) effect of patchy reionization on large scales.    Furthermore, for all mock realisations of the power spectrum the off-diagonal elements in the covariance matrix were obtained from the 80$h^{-1}$ cMpc, homogeneous-UVB Sherwood-Relics simulation with 2048$^3$ gas particles described in \citet{Puchwein2022}.  A larger box size was chosen for constructing the covariance in this work to limit any artificial correlations at log$_{10}(k/\rm km^{-1}\,s)<-2.3$.

\subsection{Neural network interpolator} \label{section:NN}

How to interpolate between different simulations to obtain a theory prediction at any point in parameter space is an issue common to all MCMC-based Bayesian inference studies of the \Lya forest. Methods currently in use are either based on linear interpolation techniques \citep{Viel2013,Irsic2017,Yeche2017,Palanque2020}, as were adopted in M22, or on alternative techniques such as Latin hypercube sampling and Gaussian process interpolation \citep[e.g.][]{Bird2019,Pedersen2021,Rogers2021,Fernandez2022}.

In this work, we propose an approach based on machine learning techniques. We train a supervised neural network to generate the 1D power spectrum from the grid of available simulations with very high levels of accuracy, making it a reliable emulator to use in parameter recovery from forthcoming high-precision data.  While a neural network trained on our set of simulations may not be immediately generalisable to simulations used in other studies, similar neural networks could very easily be trained using independent parameter grids, and transfer-learning could be invoked to avoid having to retrain the networks completely from scratch \citep{pan2009survey}.

\begin{figure*}
     \begin{subfigure}{.48\textwidth}
    \centering
    \includegraphics[trim=10 83 10 40,width=\textwidth]{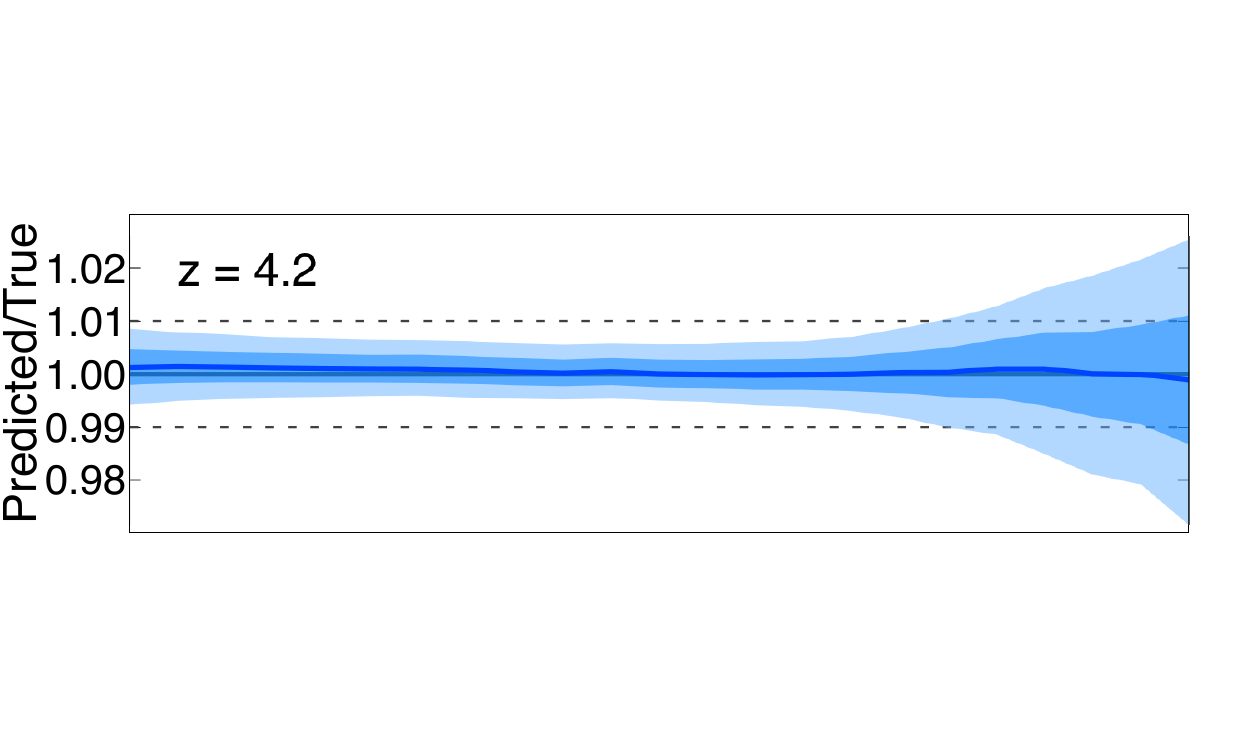} \\
    \includegraphics[trim=10 66 10 40,width=\textwidth]{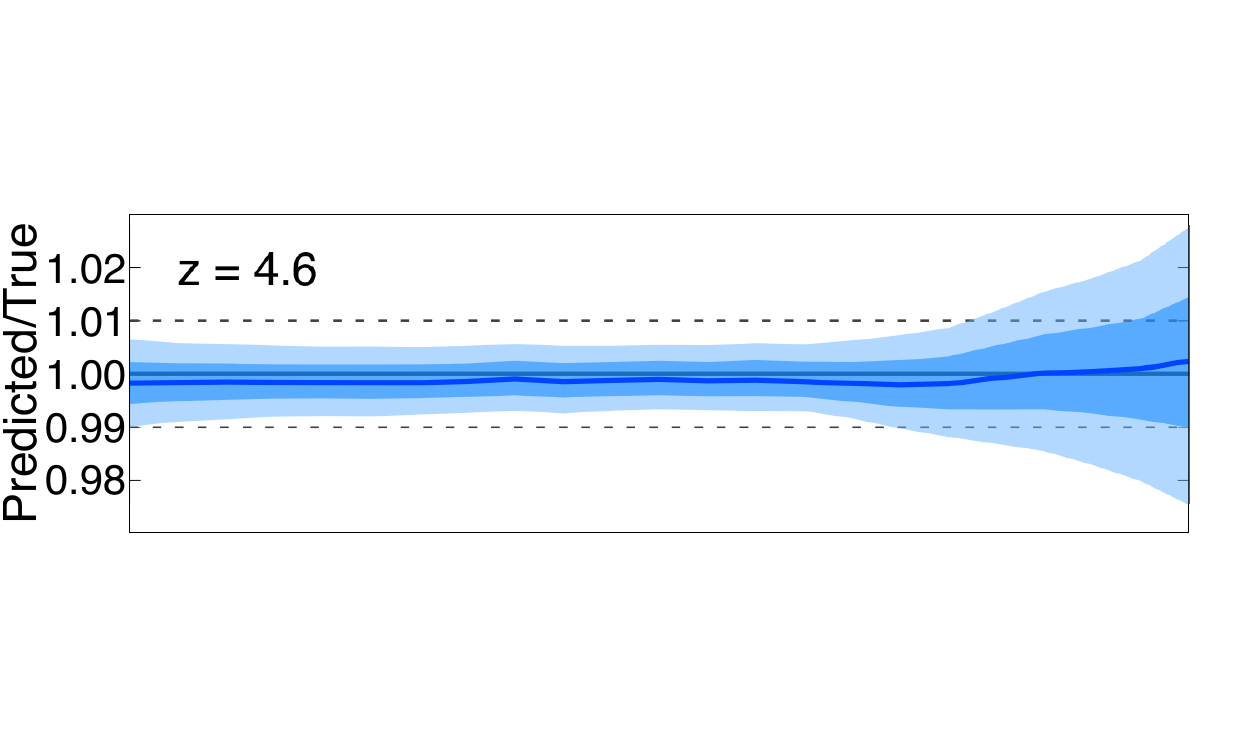} \\
    \includegraphics[trim=10 50 10 40,width=\textwidth]{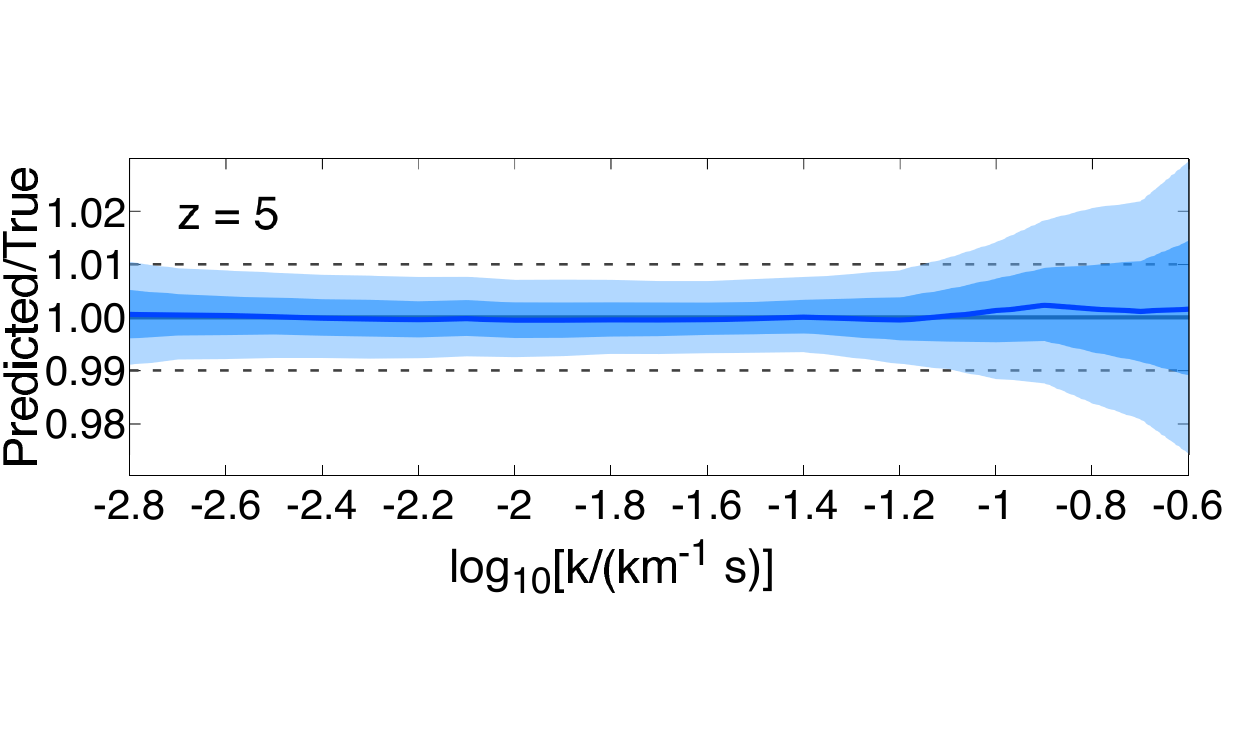} \\     
    \end{subfigure}
 \begin{subfigure}{.48\textwidth}
      \includegraphics[trim=0 40 0 0,width=\textwidth]{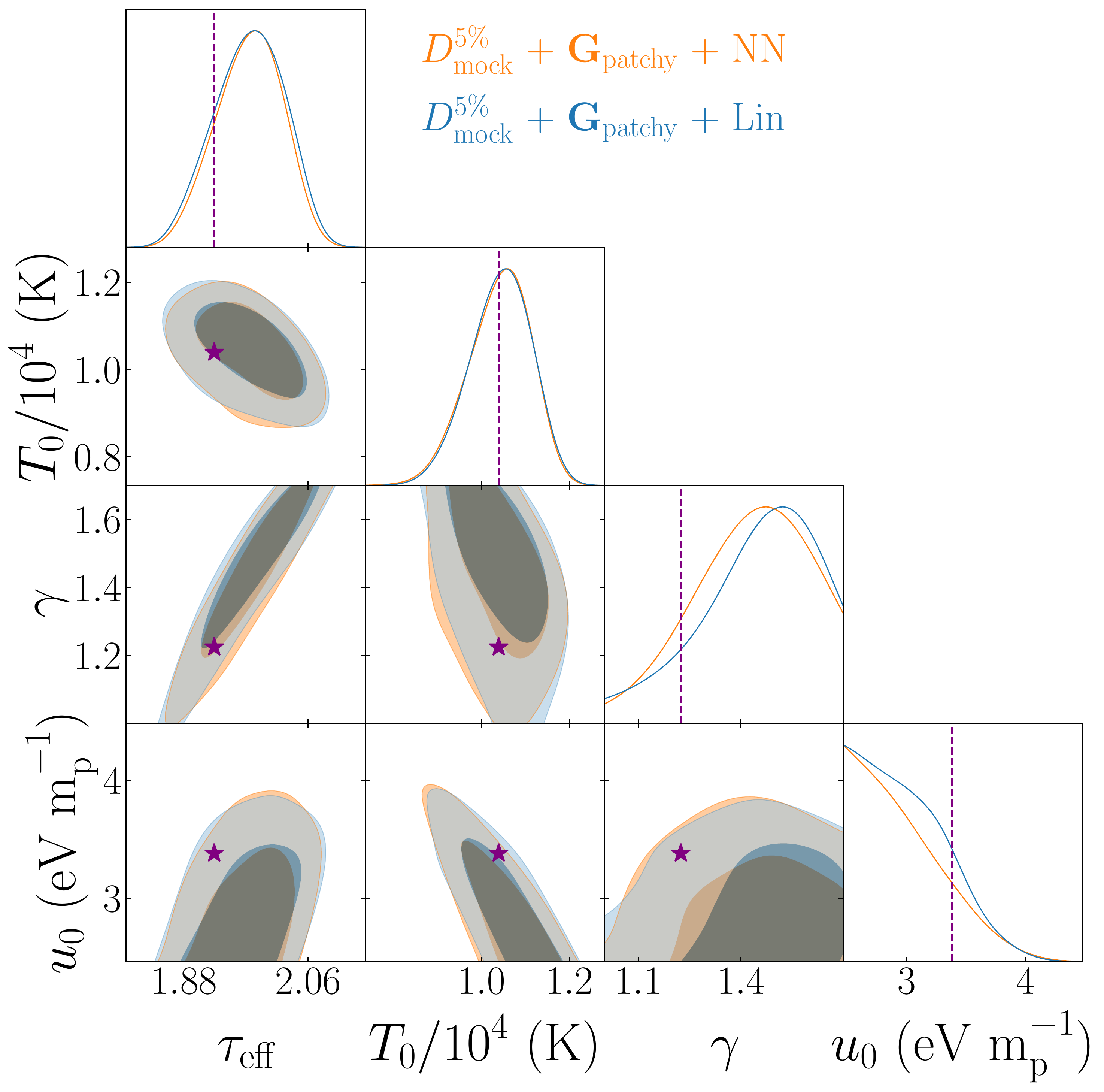} \\ 
 \end{subfigure}
    \caption{\textit{Left panel:} The distribution of the ratio between the neural-network-predicted and true power spectra in the 10-fold cross validation test. The shaded regions show the 68 and 95 per cent confidence intervals, while the solid blue line shows the median. The solid grey line indicates the perfect recovery case and the dashed lines show a $\pm$1 per cent range.  We have verified that the error distribution is well approximated by a Gaussian in all $k$ bins, and in all redshift bins. \textit{Right panel:} The one and two dimensional posterior distributions for $\tau_{\rm eff}$, $T_{0}$, $\gamma$ and $u_{0}$ at $z=5$ from our analysis of a mock \Lya forest power spectrum drawn from the RT-late simulation.   We assume 5 per cent relative uncertainties on the mock data for -2.8 $\leq$ \logtenk{} $\leq -0.7$. Results  are displayed for the linear interpolator (blue contours) or the neural-network interpolator (orange contours) using the simulation grid \Grecov{}.  The purple stars show the true parameters used in the mock data.  The bin at $z=5$ has the poorest match between the neural network and linear interpolator, which nonetheless is very good. All parameters are recovered to within 1-2 $\sigma$.}
     \label{fig:NN_testing}
\end{figure*}

The machine learning model that we adopt here is an artificial neural network -- sometimes known as a feed-forward neural network -- which consists of a series of mathematical operations applied on the input parameters (see Fig. \ref{fig:NN_diagram}). Such operations are dependent on the choice of an ``activation function'' which we denote by $\sigma$, and the particular sequence in which the operations are applied. This sequence is best described by combinations of nodes organised in layers, linked so that every node in a given layer becomes the input to every node in the following layer.  This gives rise to the "neural network" which lends its name to this method.

The output of node $m$ in layer $l$, denoted by $\textit{\textbf{a}}^{[l]}_m$, is given by
\begin{equation}
    a^{[l]}_m = \sigma(a^{[l-1]}W^{[l]\text{T}}_m + b^{[l]}_m),
\end{equation}
where the weight, \textbf{W}, and bias, \textbf{b}, are matrices whose values are iteratively constrained during the training process, and where $T$ denotes the transpose of the matrix. A neural network without an activation function is essentially a linear regression model, with the activation function adding non-linearity. The number of nodes in each layer and the total number of layers considered are free parameters that can be modified to improve the network's performance \citep{hornik1989multilayer}.

In this work, the neural network will be used to obtain predictions of the \Lya forest power spectrum at parameter values in between those already included in our simulation grid.  In our case, therefore, the input parameters $\textbf{\textit{a}}^0$ are the four astrophysical parameters that our Bayesian inference MCMC method seeks to constrain, that is \textbf{\textit{a}}$^0$ = \{$\tau\sm{eff}$, $T_0$, $\gamma$, $u_0$\}, 
where $\tau\sm{eff}$ was converted into $\langle F \rangle = \text{e}^{-\tau\sm{eff}} $ prior to training. The output of the network will be the values of the power spectrum at a discrete number, $N$, of bins in wavenumber, $k$.   As parameter recovery is performed independently using power spectrum data in separate redshift bins, we always consider independent neural networks trained, validated, and tested on 1D power spectra at the redshift under consideration.  We furthermore independently train networks on both \Ghomog{} and \Grecov{}.

In each redshift bin, the data for the training, validation, and testing of the neural network are the $18\,000$ realisations of the power spectrum  obtained from the simulations described in Section~\ref{section:grids}.   In each case, a randomly selected 90 per cent of these power spectra are used to train the neural network, while 10 per cent are saved for validation.  The loss function $L$ used for training is the mean squared error function, such that for power spectrum $t$ in the training sample, 
\begin{equation}
L_{t} = \frac{1}{N} \sum_{i=1}^{N} (\hat{P}_{k_i,t} - P_{k_i,t})^2,
\end{equation}
where $\hat{P}_{k_i,t}$ is the neural network prediction for the power spectrum, and $P_{k_i,t}$ is the true value of the power spectrum in $k$-bin $i$ for training sample $t$.  After experimenting with different activation functions we settled on the Rectified Linear Unit (ReLU) \citep{fukushima1975cognitron}, and chose a neural structure of [60,60,60]. Training was performed with batching (with batch size 12) to computationally optimise the process \citep{lecun2012efficient}.

\begin{figure*}
\centering
\begin{subfigure}{.5\textwidth}
    \centering
    \includegraphics[trim=50 0 60 0, scale=0.3]{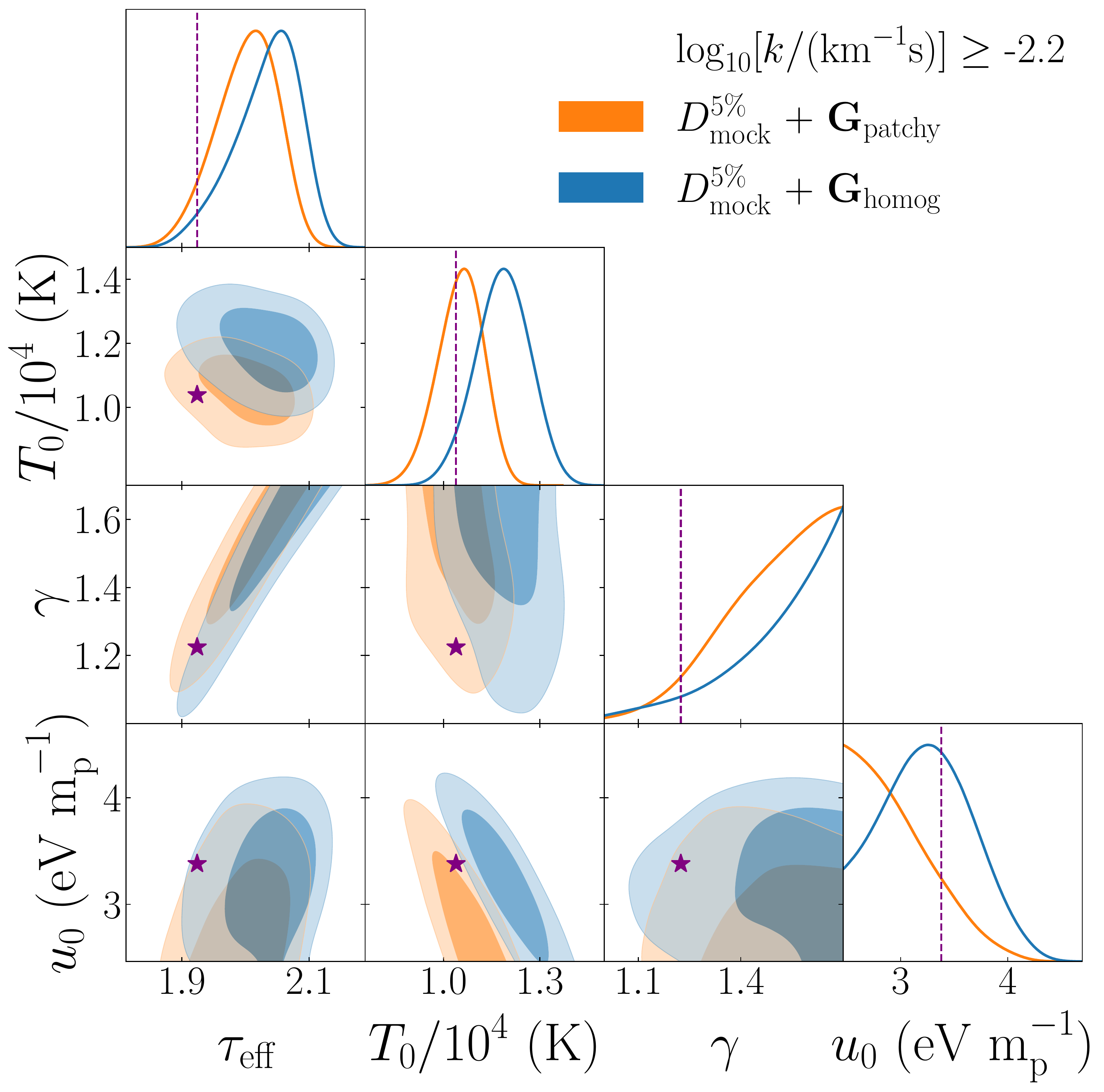}
\end{subfigure}
\begin{subfigure}{.48\textwidth}
    \centering
    \includegraphics[scale=0.3]{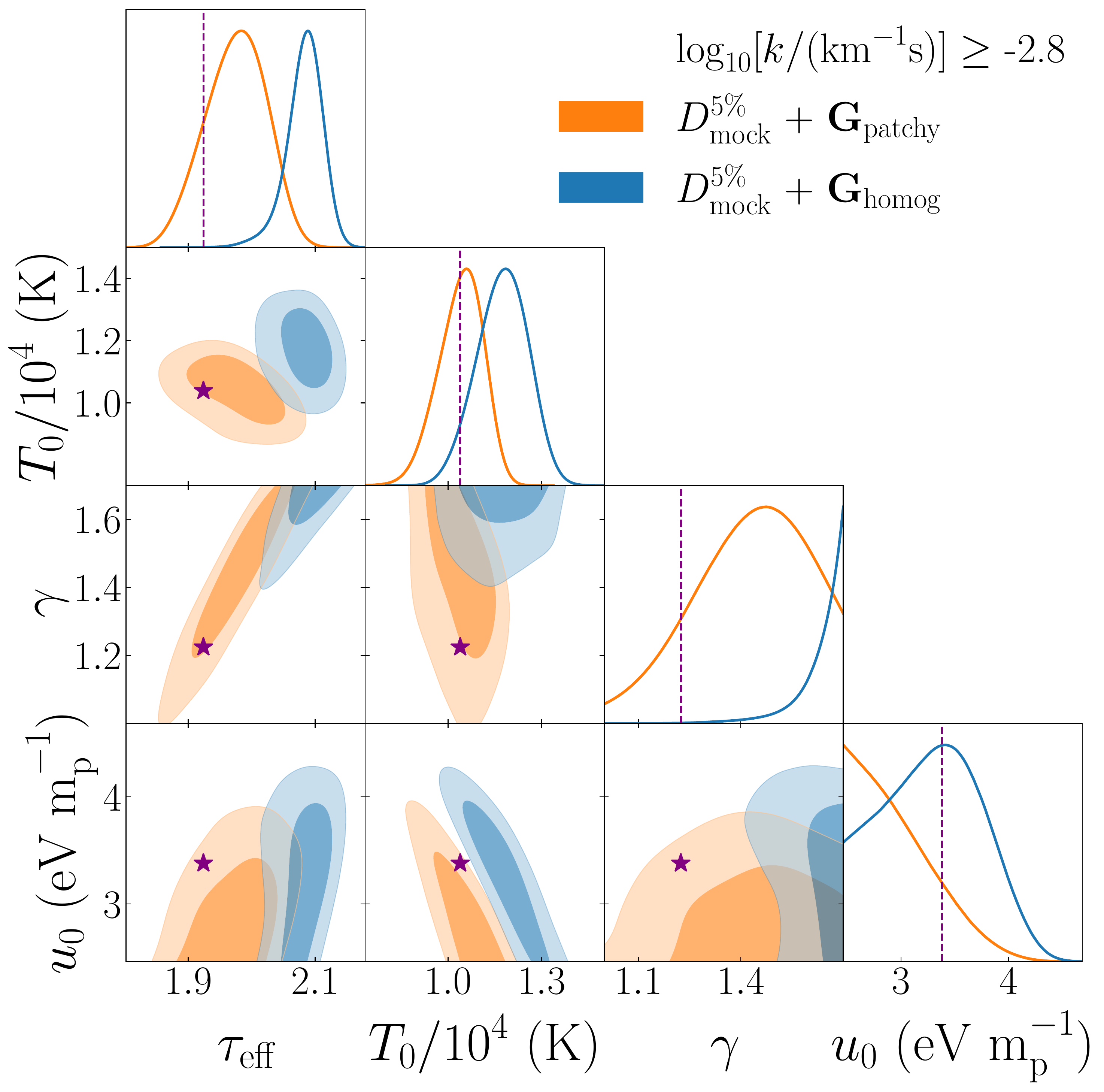}
\end{subfigure}
\vspace{-2mm}
\caption{The one and two dimensional posterior distributions for $\tau_{\rm eff}$, $T_{0}$, $\gamma$, and $u_{0}$ from an analysis of mock 1D power spectrum data drawn from the M22 RT-late simulation at $z=5$.  The mocks assume a 5 per cent relative error on the power spectrum.  The purple stars and vertical dashed lines correspond to the true parameter values in the simulation. The results are obtained using either our original grid of homogeneous UVB models (\Ghomog{}, blue contours) or that same grid including the M22 template for the effect of inhomogeneous reionization on the power spectrum (\Grecov{}, orange contours).
The left panel is for mock data spanning wavenumbers \BoeraRange{}, whereas the right panel is for mock data extending to smaller wavenumbers/larger scales, $-2.8 \leq$ \logtenk $\leq -0.7$.  Note how the parameter recovery is more sensitive to the effect of inhomogeneous reionization when including data on larger scales.  }
\label{fig:bias_increase}
\end{figure*}

The precision of our neural network was tested using k-fold cross validation \citep{stone1974cross}  as follows: the complete set of simulated realisations available in each redshift bin ($18\,000$ power spectra) was first shuffled, and then divided into $10$ sets of equal size. The initial randomization ensures that the thermal parameters of the power spectra making up each set are randomly chosen. One of the sets was then held back and the neural network was trained on the remaining $9$ sets. The neural network was then tested on the held-back set, resulting in a distribution of neural-network-predicted/true values for the $1\,800$ power spectra in the set not used in training. This process was then iteratively repeated for all the ten sets, resulting in ten independent distributions of neural-network-predicted/true values.

In Fig.~\ref{fig:NN_testing} (left panel), we show the \textit{average} distribution of the ten neural-network-predicted/true distributions independently obtained for neural networks trained on \Grecov{}.  The contour plots show (in blue) the average 68 and 95 per cent confidence intervals, while the blue solid line shows the average of the median of each distribution. The recovery error in the \logtenk{} < -1.2 range is around $0.5$ per cent at the 68 per cent confidence level, increasing up to $\sim$1 per cent at largest wavenumbers considered here (\logtenk{} = -0.6), representing a very high level of precision.  Furthermore,  we find the uncertainty in the recovery to be well approximated by a Gaussian distribution in all $k$-bins considered.

In the right panel of Fig. \ref{fig:NN_testing}, we compare the one and two dimensional posterior distributions for $\tau_{\rm eff}$, $T_{0}$, $\gamma$ and $u_{0}$ at $z=5$ from our analysis of a mock \Lya forest power spectrum drawn from the RT-late simulation described in M22.  We assume 5 per cent relative uncertainties on the mock data for -2.8 $\leq$ \logtenk{} $\leq -0.7$.   The results are obtained 
using either a linear interpolator (blue contours) or the neural-network interpolator introduced in this work (orange contours) for our \Grecov{} grid of models.  The purple stars show the true parameters of the mock data, showing that  we recover the input parameters within 1-2$\sigma$. We see that the choice of interpolator makes little difference to the parameter recovery. However, along with the significantly improved control of interpolation uncertainties, the neural network interpolator reduces the computational requirements of the MCMC sampler.  Using the same computational hardware, we find the neural network requires $\sim$ 1.5 per cent of the CPU time  required by the linear interpolator for an equal number of steps.  We will leverage the speed of this method in Section \ref{section:higher_pres}, where we perform an analysis of multiple realisations to assess the variance in our parameter recovery.


\section{The effect of inhomogeneous reionization on thermal parameter recovery}\label{section:Az_freeparam}

\subsection{Revisiting M22 using the 1D power spectrum on large scales}

\begin{figure}
\centering
\includegraphics[trim=50 0 0 0]{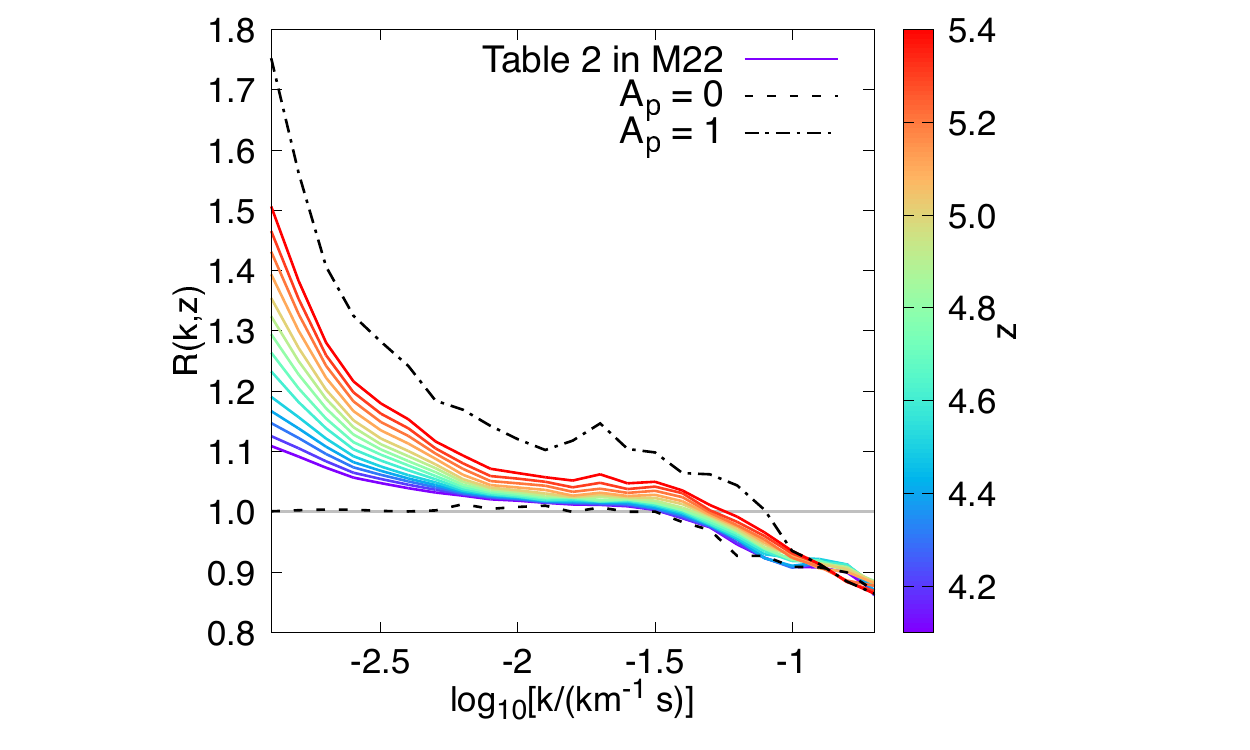}
\vspace{-0.8cm}
\caption{The coloured curves show the inhomogeneous reionization correction to the power spectrum, $R(k,z)$, presented in Table 2 of M22 (note that in this work we drop the `mid' label used by M22). The redshift at which this correction is drawn from the reionization simulation used in M22 is indicated in the colour bar.   For comparison, the dashed and dot-dashed curves show $R(k,z)$  for $A_{\rm p}=0$ and $A_{\rm p}=1$, respectively. This follows the updated parameterisation we use in this work, given by Eq.~(\ref{eqn:best_fit_free}).  The range corresponding to the tabulated M22 template is $A_{\rm p}=[0.255,0.765]$.}
\label{fig:rcorr_param}
\end{figure}

\begin{figure}
\centering
    \includegraphics[scale=0.245, trim = 110 0 110 0]{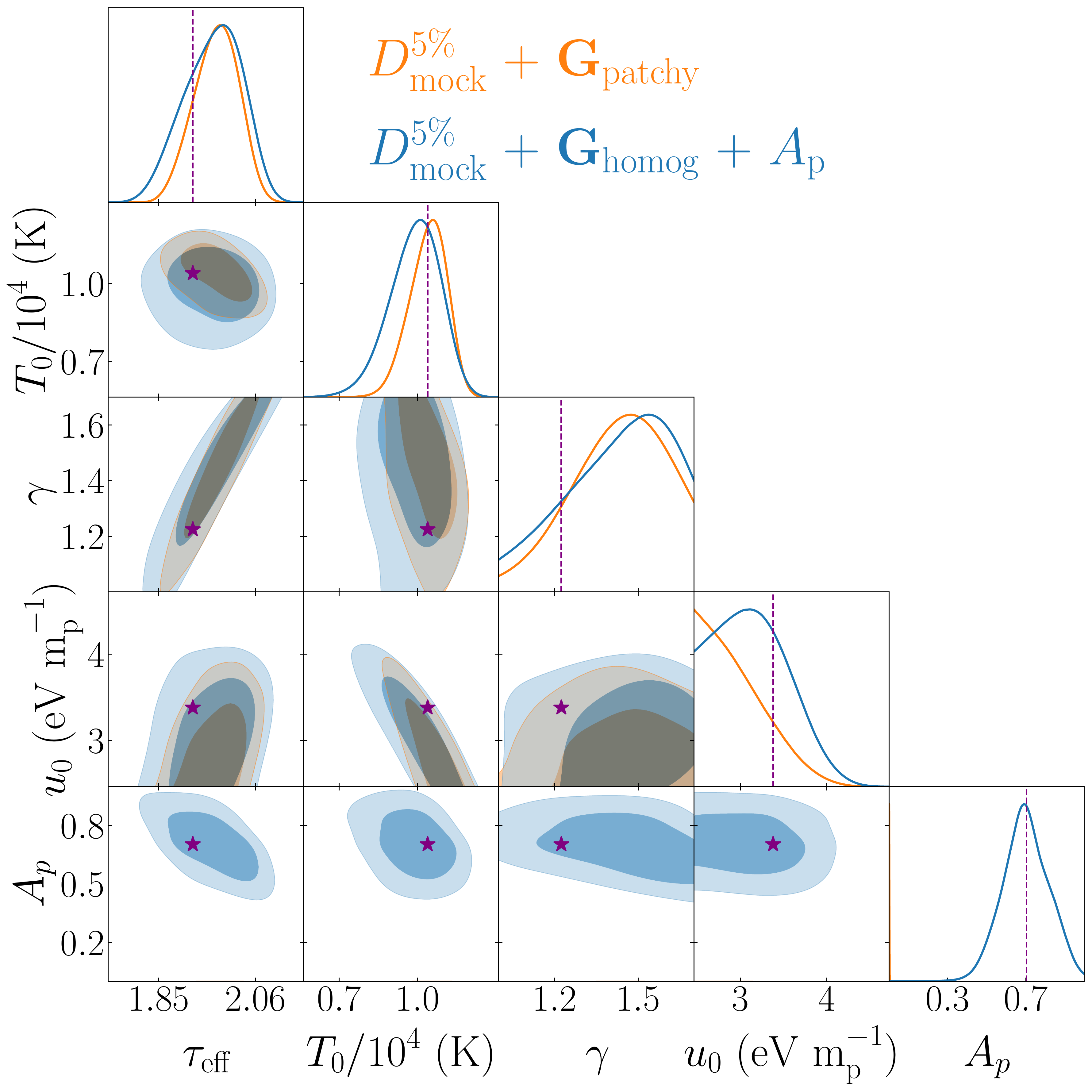}
\caption{As for the right panel of Fig.~\ref{fig:bias_increase}, but now showing the results for \Ghomog{}$+A\sm{p}$ instead of \Ghomog{}, where the former now uses the parameterisation for the patchy reionization template introduced in Eq.~(\ref{eqn:best_fit_free}) (see text in Section~\ref{sec:general} for further details).  These two approaches lead to similar parameter constraints with overlapping contours.  This indicates our updated \Ghomog{} + $A\sm{p}$ approach remains accurate, while simultaneously affording a greater degree flexibility and model independence to our analysis. }
\label{fig:mcmc_5p_panels_with_Ap}
\end{figure}

We now revisit the analysis of the 1D power spectrum presented by M22.  This earlier work concluded that the effect of inhomogeneous reionization on the shape of the power spectrum at $4.2<z<5$ should not strongly bias \emph{existing} constraints on IGM thermal parameters obtained from an analysis of high resolution \Lya forest data \citep[see also][]{Wu2019}.  This conclusion was based on assuming a relative uncertainty of 10 per cent on power spectrum measurements (i.e. similar or smaller than the level of precision currently achieved for the high resolution data presented by \citet{Boera2019}).  On the other hand, M22 noted that, for data with relative uncertainties at the 5 per cent level, patchy reionization effects that change the shape of the power spectrum at small scales, $\log_{10}(k/{\rm km^{-1}\,s})>-1.5$  will introduce a modest ($\sim 1 \sigma$) shift in the recovery of IGM thermal parameters.  M22 found this was primarily driven by divergent peculiar velocity gradients and variations in the thermal broadening kernel that alter the shape of the power spectrum on small scales.

However, the power spectrum analysis presented by M22 only considered wavenumbers $\text{log}_{10}[k/(\text{km}^{-1} \text{s})] \geq -2.2$, matching the measurement range presented by \citet{Boera2019}.  As discussed earlier, this excludes the larger scales where the  enhancement in the power spectrum due to relic fluctuations in the neutral hydrogen fraction from reionization are expected to be largest.  In this context, surveys such as DESI \citep{VargasMagana2019} and WEAVE-QSO \citep{Pieri2016} will not only achieve higher levels of precision, but will also probe larger physical scales in the redshift range $2 \leq  z \leq  4.5$, where enhancements to the 1D power spectrum arising from patchy reionization are expected to be most prominent.  Here, we therefore quantify how patchy reionization will impact on the recovery of IGM parameters by applying the M22 analysis to smaller wavenumbers/larger scales where the patchy correction is largest.

In Fig. \ref{fig:bias_increase}, we use our parameter estimation framework to consider parameter recovery from mock data drawn from the RT-late simulation used in M22 \citep[see also][]{Puchwein2022}.  This simulation follows inhomogeneous reionization ending at redshift $z\sm{R} = 5.3$, where we choose to define the end point of reionization as the redshift when the volume-averaged \HI fraction first falls below $x\sm{HI} = 10^{-3}$ in the model.  We assume relative uncertainties of 5 per cent on the power spectrum and consider two different wavenumber ranges: $-2.2 \leq \log_{10} (k/{\rm km^{-1}\,s})\leq -0.7$ in the left panel (i.e. the range considered by M22 and \citet{Boera2019}), and a range that extends to larger scales, $-2.8 \leq \log_{10} (k/{\rm km^{-1}\,s})\leq -0.7$, in the right panel.

The panels in Fig.~\ref{fig:bias_increase} show the one and two dimensional posterior distributions recovered for $\tau_{\rm eff}$, $T_{0}$, $\gamma$ and $u_{0}$ in a single redshift bin at $z=5$.  The results are obtained using our neural network power spectrum emulator, for either the grid of homogeneous UV background models used in M22 (\Ghomog{}, blue contours), or that same grid including a template that captures the effect of inhomogeneous reionization on the power spectrum, based on eq. (3) in M22 (\Grecov{}, orange contours). The purple stars and vertical dashed lines correspond to the true parameter values used in the simulation.

First, as already noted in M22 (see their fig. 10), for wavenumbers $-2.2 \leq \log_{10} (k/{\rm km^{-1}\,s})\leq -0.7$ patchy reionization introduces a modest $\sim 1\sigma$ shift between the true and recovered parameters (i.e. the blue contours and purple stars in the left panel of Fig.~\ref{fig:bias_increase}) when assuming a uniform UV background (\Ghomog{}).  However, when including larger scales in the right panel, the systematic bias in the parameter recovery is much larger and is at the level of $\sim 6 \sigma$ for the total $\chi^{2}$.  The largest shift is for $\tau\sm{eff}$; this is unsurprising, given that $\tau\sm{eff}$ sets the amplitude of the power spectrum on the scales where the patchy reionization effects are largest. There is also a large shift in the temperature-density relation, however this is entirely due to the very strong degeneracy between $\tau_{\rm eff}$ and $\gamma$.  We also observe that the simulation parameters are generally well recovered (within $\sim 1$--$2\sigma$) when applying our patchy reionization template to the grid of homogeneous UV background models (\Grecov{}), even when extending our analysis to larger scales. This is expected and serves as a useful consistency test, given that these mock data were drawn from one of the simulations used in M22 to obtain the patchy reionization template.

Parameter recovery is therefore much more sensitive to the effect of inhomogeneous reionization when including data on larger scales.   As has been noted elsewhere  \citep[e.g.][]{Cen2009,Keating2018,DAloisio2018fluc,Onorbe2019,Montero2020}, this raises the interesting possibility that -- given a significant detection of enhanced large scale power --  the end stages of reionization may be constrained with precision measurements of the 1D \Lya forest power spectrum on scales $\log_{10}(k/{\rm km^{-1}\,s})\simeq -3$ at $z>4$.

\subsection{A generalised approach to modelling the enhanced large scale power from patchy reionization} \label{sec:general}

Before turning to consider observational data, we additionally modify the framework first introduced by M22 to provide a more flexible, model independent approach for quantifying the amount of excess large scale power in data.  The motivation for this is two-fold. First, it avoids using a parmeterisation that is directly tied to the redshift evolution of the reionization models used in M22.  Second, it allows for a general parametrisation of the excess of large-scale power that can capture a larger variety of models. In the same manner as cosmological parameter data compression \citep[e.g.][]{McDonald2000,Pedersen2022}, the interpretation of the measured excess large scale power can then be mapped back onto physical models.

We implement this updated approach by continuing to use the template provided in Table 2 of M22, where the shape of the correction to the power spectrum for patchy reionization  -- obtained from radiative transfer simulations  -- is given by the quantity $R(k,z)$.  Here
\begin{equation} R(k,z)=\frac{P_{\rm RT}(k,z)}{P_{\rm homog}(k,z)}, \end{equation}
where $P_{\rm RT}(k,z)$ ($P_{\rm homog}(k,z)$) represent the 1D flux power spectrum from simulations that include (ignore) the effect of inhomogeneous reionization on the \Lya forest, but otherwise have the same \emph{volume-averaged} reionization history \citep[see][for further details]{Puchwein2022}.  However, instead of assuming a model dependent redshift evolution for this template that is tied to the adopted reionization history (i.e. eq. (3) in M22), we now allow $R(k,z)$ to \emph{vary freely} across each redshift bin within our analysis.  We implement this by introducing the free parameter $A\sm{p}$ with a flat prior in the range $[0,1]$.  This parameter controls the amplitude of the imprint of patchy reionization on the 1D flux power spectrum. This is achieved by relating it to the redshift, $z$, at which the correction is linearly interpolated from the data in Table 2 of M22 by 
\begin{equation}
\label{eqn:best_fit_free}
    z = 3.45 + 2.55 A\sm{p},\quad \textnormal{for}\,0\leq A_{\rm p} \leq 1.
\end{equation} 
This linear mapping has been obtained using the patchy reionization template $R(k,z)$ at $4.1\leq z \leq 5.4$ in table 2 of M22, and linearly extrapolating to a lower and upper redshift limit of $z=3.45$ and $z=6$, respectively. The lower redshift is chosen such that the extrapolated template correction is negligible on large scales, which we find occurs at $z=3.45$ (i.e. for $A_{\rm p}=0$).  The extrapolation similarly gives $A_{\rm p}=1$ by $z = 6$.  In practice, however, the exact parameterisation makes little difference to our results, as our main aim is to establish if $A_{\rm p}\neq 0$ is preferred in observational data.  This is illustrated further in Fig. \ref{fig:rcorr_param}, where the coloured curves show the data presented in Table 2 of M22, and the dashed (dot-dashed) curves correspond to $A_{\rm p}=0$ ($A_{\rm p}=1$) within our new parameterisation.  We will refer to this approach as  \Ghomog$+A\sm{p}$ within our analysis framework.

In Fig. \ref{fig:mcmc_5p_panels_with_Ap} we perform a brief consistency test of this approach, by once again examining the one and two dimensional posterior distributions obtained from the mock data used in Fig.~\ref{fig:bias_increase}.  Recall the mock data are drawn from the RT-late simulation in M22 at $z=5$, and assume a 5 per cent relative error on the power spectrum measurement.   The orange contours are identical to those shown in Fig.~\ref{fig:bias_increase} (\Grecov{}), while the blue contours show the results for the generalised \Ghomog{}$+A\sm{p}$ approach. The purple stars and vertical lines, as usual, represent the true parameters in the mock data.  Note that \Grecov{} corresponds to a fixed value of $A_{\rm p}$ (as obtained by comparing Eq.~(\ref{eqn:best_fit_free}) to eq.~(3) of M22), such that $R(k,z)$ is assumed to exactly match the shape expected for a model with reionization ending at $z_{\rm R}=5.3$, as is appropriate for the model from which the mock data was produced.  We observe that the two sets of contours overlap and $A_{\rm p}$ is recovered within $1\sigma$, indicating this approach maintains the accuracy of our parameter recovery, while also affording a greater degree of generality.


\section{A hint of enhanced large scale power in the flux power spectrum} \label{section:karac_data}

\subsection{Observational data}

We now apply our updated analysis framework to the observational measurements of the 1D \Lya forest power spectrum at $4 \leq z \leq 4.6$ presented recently by \cite{Karacayli2022}.  The \cite{Karacayli2022} flux power spectrum has been obtained using a collection of high resolution and high signal-to-noise data from VLT/XSHOOTER (XQ-100, \citet{Lopez2016}), VLT/UVES (SQUAD, \citet{Murphy2019}) and Keck/HIRES (KODIAQ, \citet{OMeara2017}). The flux power spectrum measurements are made in a total of $15$ redshift bins, in the range of $z=1.8-4.6$ with bin size $\Delta z=0.2$. This work focuses on high redshifts only ($z>4$), because the relic fluctuations of reionization dissipate at lower redshifts (see M22 and Fig.~\ref{fig:rcorr_param}).  In each of the redshift bins, the flux power is measured in $21$ $k$-bins, from $k=0.0011\;\skm$ to $k=0.39569\;\skm$ using non-uniform spacing (the bins are equidistant in $k$ on large scales, and equidistant in $\log_{10}{k}$ on small scales, with the divide at $k\sim 0.01\;\skm$). Following the discussion of \citet{Karacayli2022} on the effects of noise and resolution on the smallest scales, and because we are interested in isolating the effect of inhomogeneous reionization on large scales, we further restrict our analysis by limiting the wavenumber range to $k<0.1\;\skm$. This leaves $15$ $k$-bins in each of the 4 redshift bins, for a total of 60 data points.

By virtue of combining several data sets, these measurements boast one of the highest precision measurements to date of the flux power spectrum using high resolution Lyman-$\alpha$ forest data. A typical uncertainty on the flux power spectrum at $z=4.2\;(4.6)$ is at the level of $\sim 10-15\text{ per cent}\;(20-30\text{ per cent})$, ranging from intermediate to large scales. This is similar to the less sparsely sampled measurements of \citet{Boera2019}, who use bin sizes $\Delta z=0.4$ and report uncertainty levels on the flux power spectrum of $\sim 10-20\text{ per cent}\;(8-10\text{ per cent})$ at $z=4.2$ (4.6). Furthermore, the \citet{Karacayli2022} power spectrum analysis is performed jointly on the quasar spectra from all three samples using the optimal quadratic estimator \citep{Karacayli2020}.  An advantage over the more common fast Fourier transform estimators found in the literature \citep[e.g.][]{Viel2013,Irsic2017_XQ100,Walther2019} lies in better control of the masked regions and therefore better recovery of the large-scale modes. This is critical for our analysis, as the effect of inhomogeneous reionization on the power spectrum is greatest on the largest scales probed by the data.  Note also that although the flux power spectrum measurements from large spectroscopic surveys such as the extended Baryon Oscillation Spectroscopic Survey (eBOSS) reach a higher statistical precision \citep[e.g.][]{Chabanier2019}, the typical spectral resolution  prevents these surveys from measuring small-scale power.  At large scales, the IGM thermal parameters are highly degenerate with the additional enhancement of power from reionization fluctuations, and adding small-scale information can significantly help in breaking this degeneracy \citep[e.g.][]{Viel2013,Irsic2017,Yeche2017}.  \edit{Furthermore, different systematic uncertainties can impact on the power spectrum measurements from low and high resolution data (see Section~\ref{sec:alt}), such that a joint analysis with eBOSS data would be more complicated.  For these reasons, we limit our investigation to the high-resolution data of \citet{Karacayli2022} and leave a joint analysis with low-resolution data to future work.}

\subsection{Data analysis}

\begin{figure*}
\centering
\begin{subfigure}{.45\textwidth}
    \centering
    \includegraphics[scale=0.245, trim = 140 0 0 0 0]{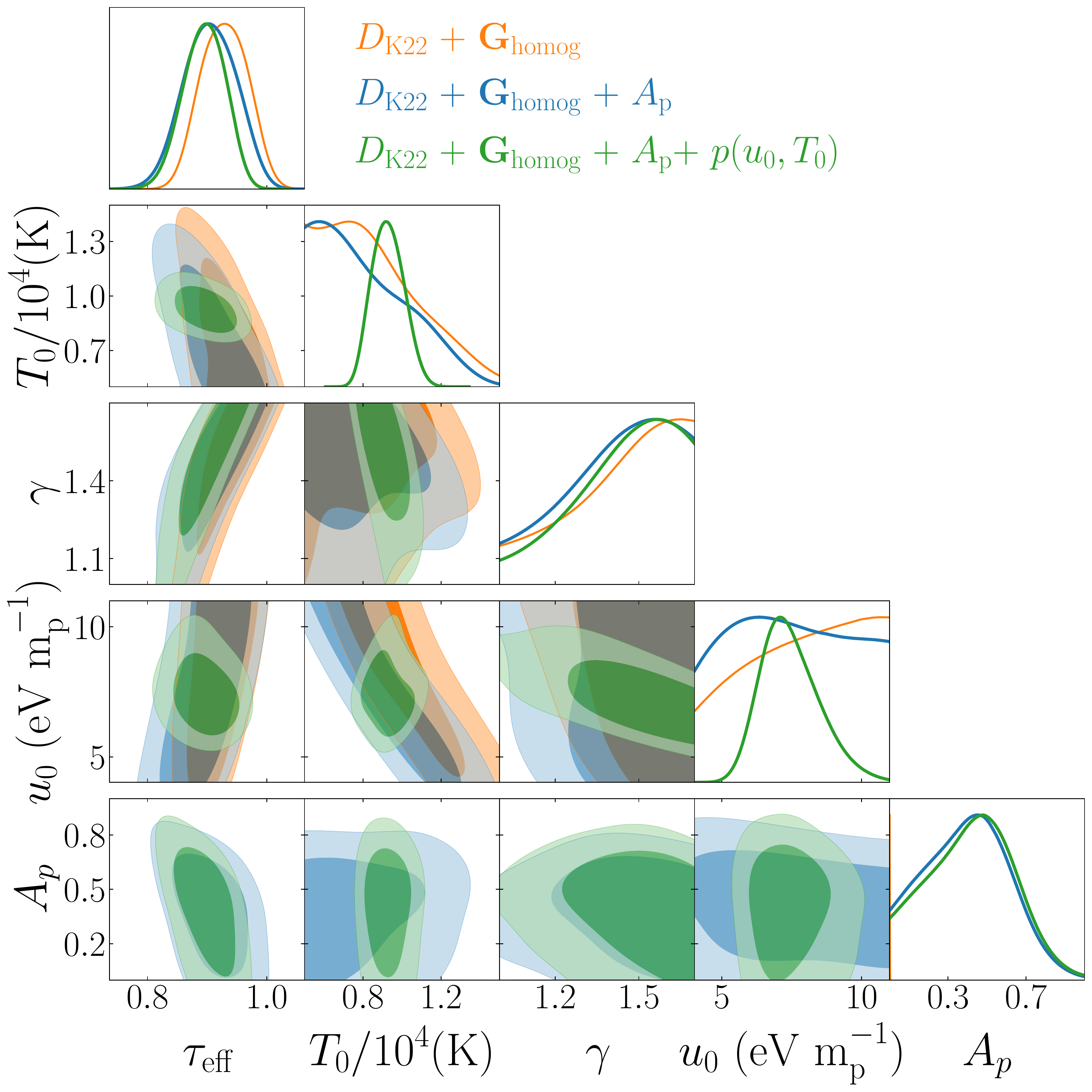}
    \caption{$z=4.0$}
\end{subfigure}%
\begin{subfigure}{.45\textwidth}
    \centering
    \includegraphics[scale=0.245, trim = 0 0 0 0 140]{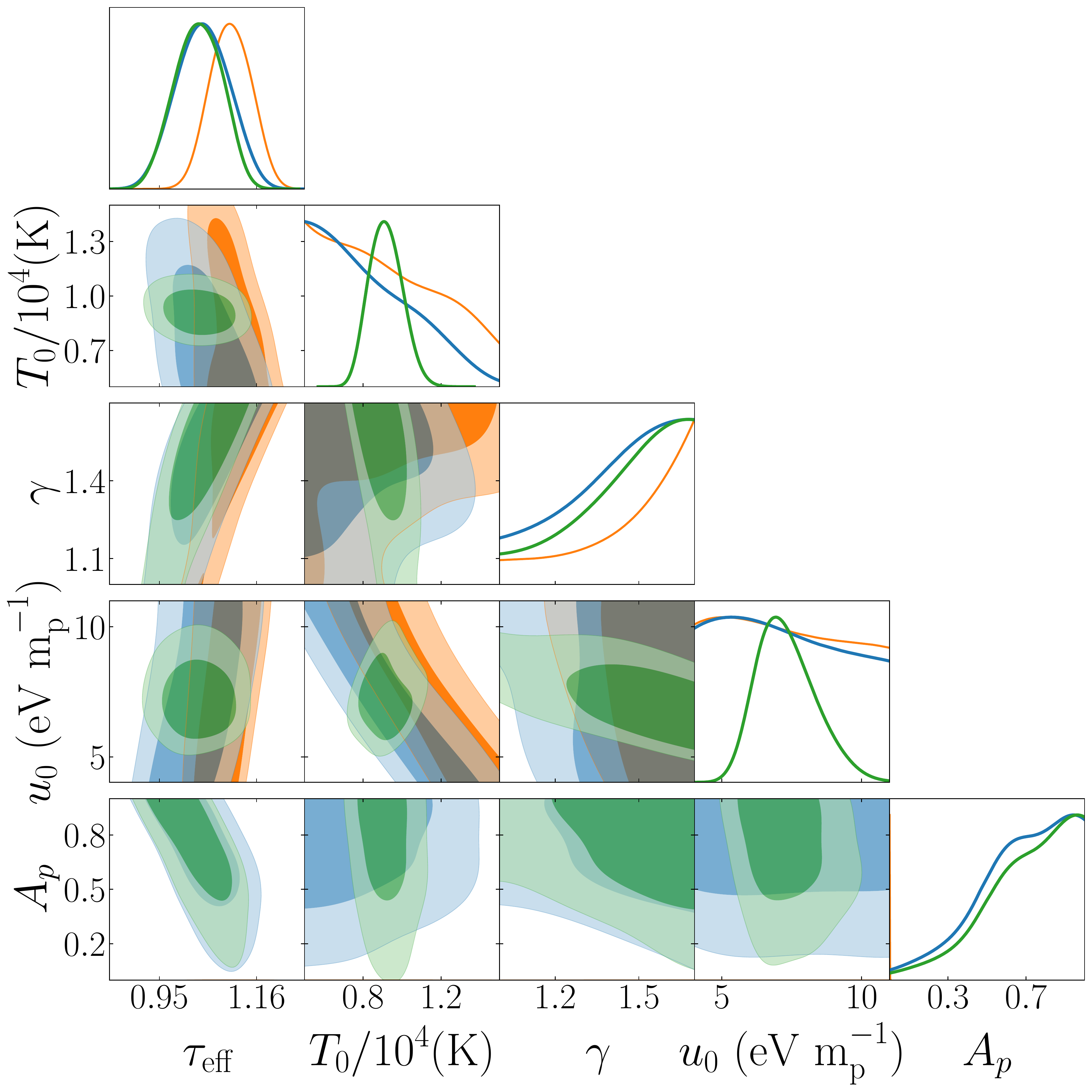}
     \caption{$z=4.2$}
\end{subfigure}%
\vskip\baselineskip
\centering
\begin{subfigure}{.45\textwidth}
    \centering
    \includegraphics[scale=0.245, trim = 140 0 0 0 0]{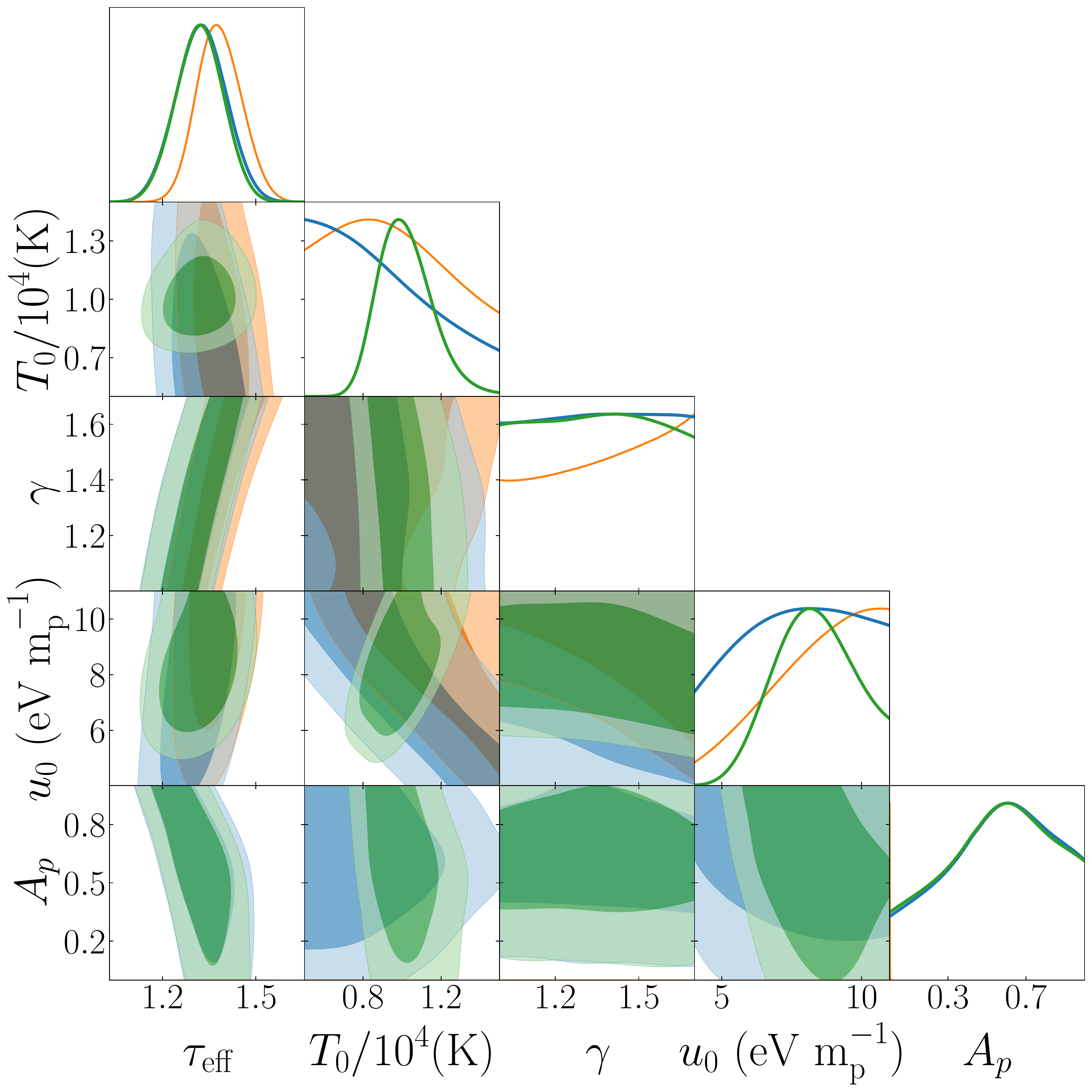}
     \caption{$z=4.4$}
\end{subfigure}%
\begin{subfigure}{.45\textwidth}
    \centering
    \includegraphics[scale=0.245, trim = 0 0 0 0 140]{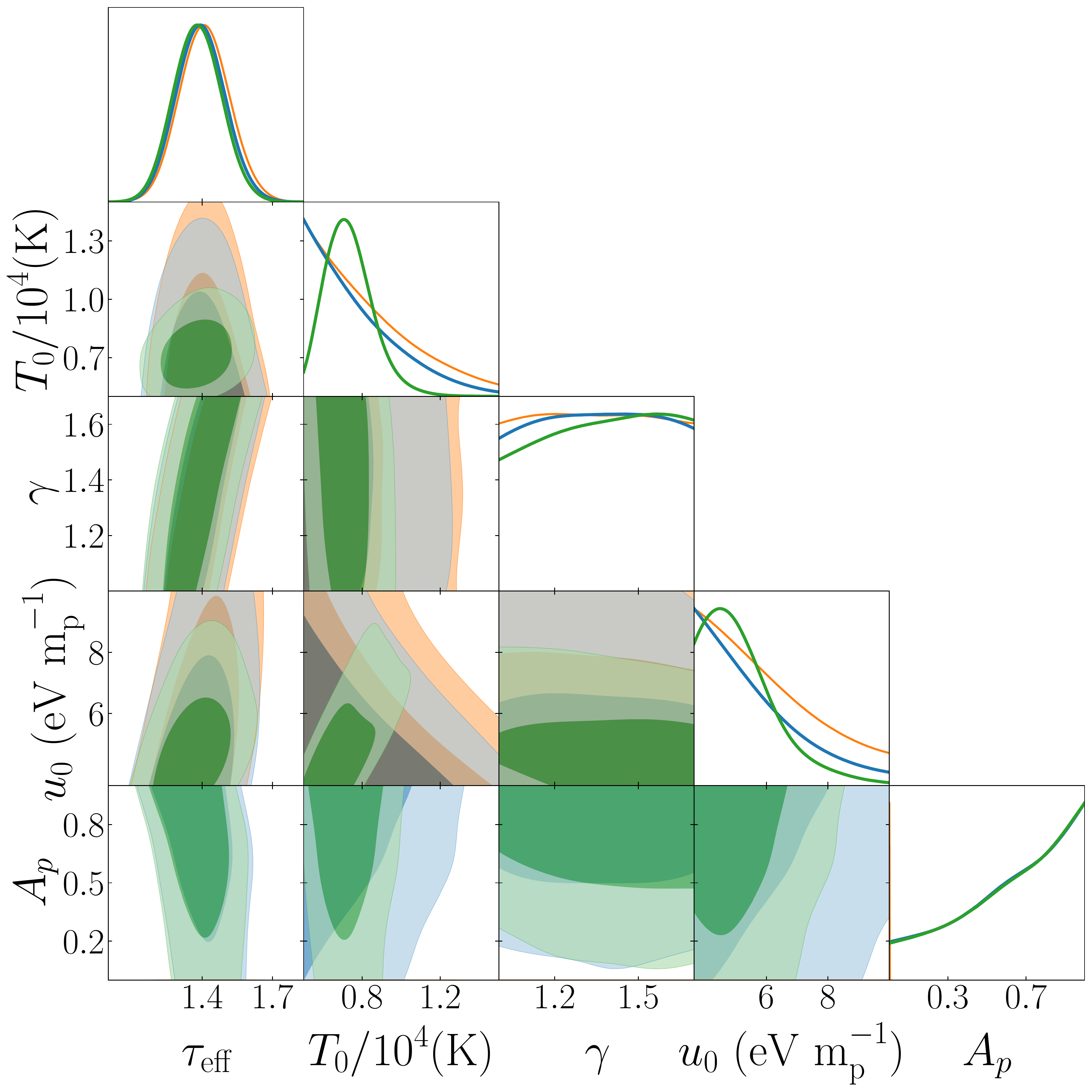}
     \caption{$z=4.6$}
\end{subfigure}%
\vspace{-1mm}
\caption{The one and two dimensional posterior distributions for $\tau_{\rm eff}$, $T_{0}$, $\gamma$, $u_{0}$ and $A_{\rm p}$ from the analysis of the \citet{Karacayli2022} data using \Ghomog{} (orange contours) and \Ghomog{}$+A_{\rm p}$  (blue contours).  The shaded regions show the 68 and 95 per cent confidence intervals for the joint distribution.   For the latter, we also show in green the contours obtained when including the $u_0 - T_0$ prior described in Section \ref{section:karac_data}. The four panels are for the four \citet{Karacayli2022} redshift bins analysed in this study, namely $z = 4.0,4.2,4.4, 4.6$.}
\label{fig:mcmc_karac_data}
\end{figure*}

\begin{figure*}
    \centering
    \begin{subfigure}{0.3\textwidth}
     \includegraphics[trim = 150 30 80 0]{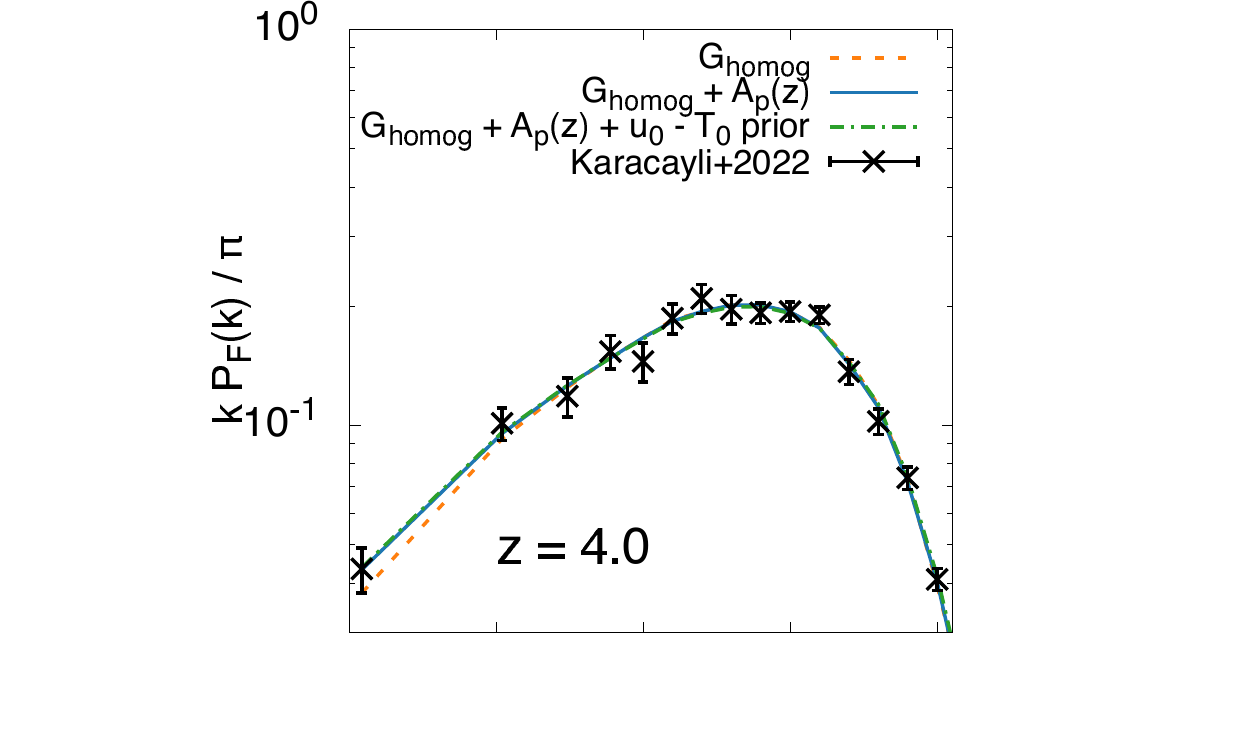}   
     \includegraphics[trim = 150 0 80 0]{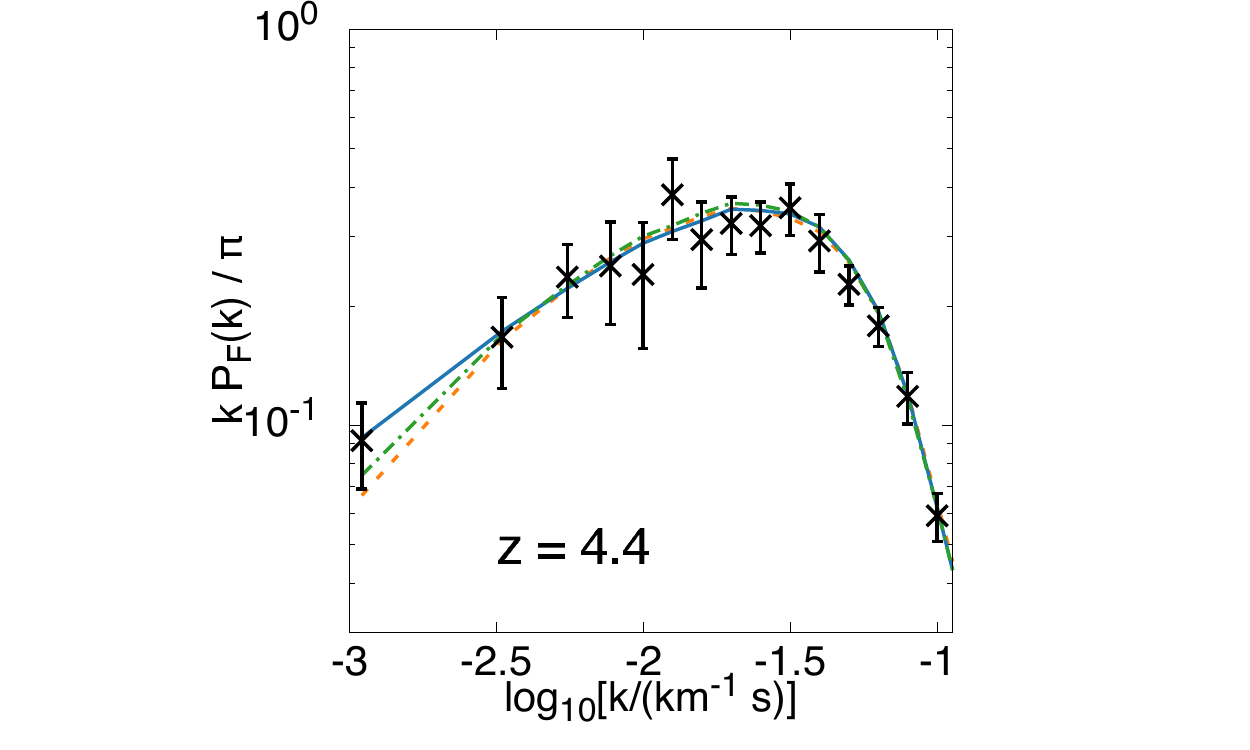}    
    \end{subfigure}
    \begin{subfigure}{0.3\textwidth}
     \includegraphics[trim = 80 30 170 20]{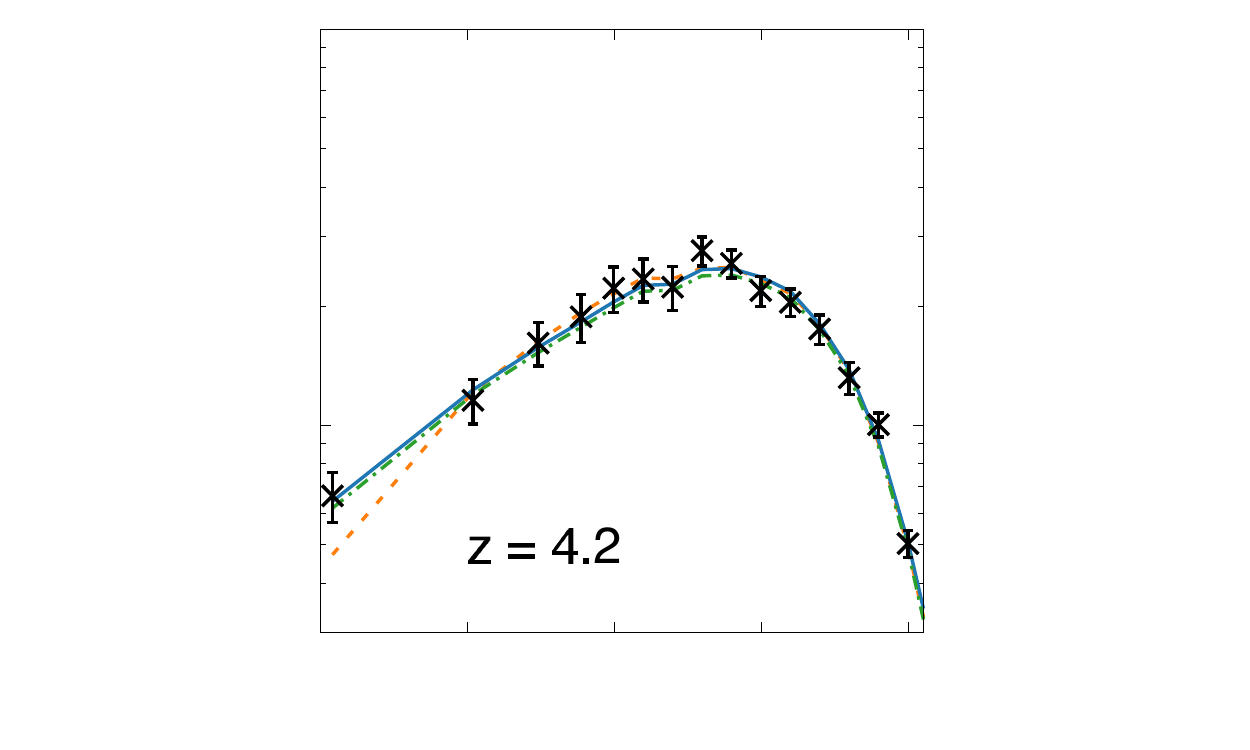}   
         \includegraphics[trim = 80 0 170 0]{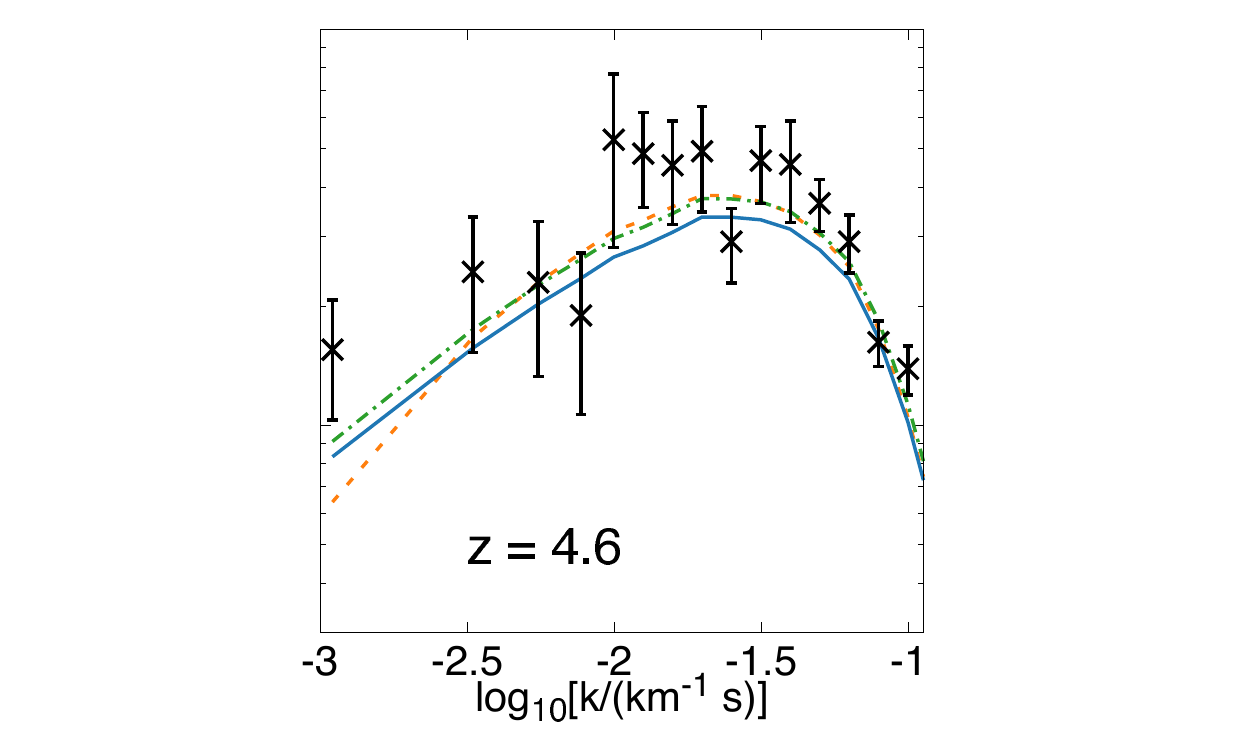}
    \end{subfigure}
    \label{fig:my_label}
    \vspace{-4mm}
    \caption{The best-fit $P_{\rm F}(k)$ models obtained from our MCMC analysis of the \citet{Karacayli2022} data for the three cases displayed in Fig. \ref{fig:mcmc_karac_data}: \Ghomog{} shown by the orange dashed curves, \Ghomog + \Ap{} shown by the solid blue curves, and \Ghomog + \Ap{} + $u_0-T_0$ prior shown by the dotted-dashed green curves. The data points with error bars show the \citet{Karacayli2022} power spectrum measurements in the four redshift bins considered here, $z = 4.0,4.2,4.4, 4.6$. The best-fit $\chi^2$/d.o.f. values for the three cases are 55.4/44, 41.6/40, and 43.3/40 respectively, corresponding to p-values of $p=0.12$, $0.40$ and $0.33$.  The data therefore exhibit a preference for an enhancement in the large scale power relative to models that assume a spatially homogeneous UV background.}
    \label{fig:karac_with_bestfit}
\end{figure*}

In order to compare our models to the flux power spectrum measurements of \citet{Karacayli2022}, we use the same strategy and grid of simulations described in Section~\ref{section:emulator} and in M22.  However, we make two minor but important modifications.  First, because we are comparing our models to observational data, we must apply a mass resolution correction to all of our simulated models, which have been performed in $20h^{-1}\rm\,cMpc$ boxes with a gas particle mass of $M_{\rm gas}=9.97\times10^{4}h^{-1}\,M_{\odot}$ (see Table 1 in M22).  The resolution correction is obtained using a grid of simulations that have parameters matched to those used in M22, but with a mass resolution that is improved by a factor of $8$.\footnote{Specifically, we use simulations performed with box sizes of $10h^{-1}\rm\,cMpc$ using $2\times 1024^{3}$ particles, yielding a gas particle mass of $M_{\rm gas}=1.25\times 10^{4}h^{-1}\,M_{\odot}$  \citep[see Table 1 of][for full details]{Puchwein2022}.}  This  correction is largest at small scales, and is at most 12 per cent at $k=0.1\rm\,km^{-1}\,s$ ($z=5$), but is very small ($<1$ per cent) on the large scales where the effect of patchy reionization is largest. Secondly, we now train the neural network on post-processed power spectra that use the same $k$-binning as the \cite{Karacayli2022} data.   We have verified that the accuracy and precision of the trained neural network remains at the same level demonstrated in Fig.~\ref{fig:NN_testing} earlier.

In Fig.~\ref{fig:mcmc_karac_data} we show the one and two dimensional posterior distributions obtained from the analysis of the \cite{Karacayli2022} data excluding (\Ghomog{}, orange contours) or including (\Ghomog{} + \Ap{}, blue contours) our parameterisation for the effect of inhomogeneous reionization on large scales.   For the latter, we also consider a third case where a non-flat $u_0 - T_0$ prior is added to the analysis.  We add this because the posterior on $u_0-T_0$ is wide and some regions of parameter space (e.g. where $T_{0}$ is very large and $u_{0}$ is very low) are unphysical.  Hence, instead of allowing $u_{0}$ and $T_{0}$ to vary freely, we apply a prior that encompasses a physically plausible region within the $u_0-T_0$ plane.  We base this on the thermal histories used in our hydrodynamical simulations, corresponding to a region bounded by $6 \leq (u_0/{\rm eV\,m_{\rm p}^{-1}}) (T_{0}/10^{4}\rm\,K)^{-1.7} \leq 12$ and $0.5\leq T_{0}/10^{4}\rm\,K \leq 1.5$.   This prior enforces a tighter correlation between $u_0-T_0$ that is almost perpendicular to the degeneracy axis between $u_0$ and $T_0$ found in the data.  We note, however, that the $u_0 - T_0$ prior only affects parameter recovery from the power spectrum at the smallest scales, $k\sim 0.1\rm\,km^{-1}\,s^{-1}$, that are most sensitive to pressure smoothing and the thermal broadening kernel \citep{Nasir2016}; there is little effect on the recovery of \Ap{} as a consequence. This result further suggests that the patchy reionization information contained in the \Ap{} parameter is sensitive to the flux power spectrum on large scales only.

Finally, in Fig. \ref{fig:karac_with_bestfit}, we compare the best-fit power spectrum models obtained from this analysis to the \cite{Karacayli2022} data in each redshift bin.  Note that the goodness-of-fit varies across individual redshift bins, with the poorest at $z=4.6$ where -- although the error bars are generally larger -- there is some tension between the models and the change of the amplitude of the observed power spectrum from small to large scales. Note also that adjacent redshift bins are correlated, such that performing a simple ``chi-by-eye'' can be misleading.  The joint fit across all four redshift bins is reasonable, with  best-fit $\chi^2$/d.o.f. values of 55.4/44, 41.6/40, and 43.3/40 for the \Ghomog{}, \Ghomog{} + \Ap{}, and \Ghomog{} + \Ap{} + $u_0 - T_0$ prior cases respectively, corresponding to p-values of $p=0.12$, $0.40$ and $0.33$.   The improved $\chi^2$/d.o.f. values for the \Ghomog{} + \Ap{} cases with or without the $u_0 - T_0$ prior indicate that introducing the parameter \Ap{} in the MCMC analysis -- which introduces a boost to the large scale 1D power spectrum and a small suppression at small scales (see Fig.~\ref{fig:rcorr_param} and M22) -- leads to a better fit to the observational data.    There is a preference for \Ap{} being required by the data at a significance of $2.7\sigma$ (i.e. for an enhancement in the 1D power spectrum at large scales, $\log(k/\rm\,km^{-1}\,s)\sim -3$, relative to \Lya forest models that assume a homogeneous UV background at $z>4$). This is in qualitative agreement with the independent analysis presented by \citet{DAloisio2018fluc}, although the formal significance we derive is higher. This may hint at the presence of relic fluctuations in the ionization and thermal state of the IGM following the completion of reionization \citep[e.g.][]{Cen2009,Keating2018,Onorbe2019,Wu2019,Montero2020,Puchwein2022}.

\subsection{Alternative explanations for enhanced large scale power} \label{sec:alt}

Although the preference for non-zero \Ap{} may be a signature of patchy reionization, other factors may be contributing to the boost in the 1D power spectrum at large physical scales.  We briefly discuss some other possibilities here.

First, the enhanced flux power spectrum at large scales could have a non-astrophysical origin.  Estimators used for the statistical analysis of fluctuations in the transmitted \Lya flux require modelling of the intrinsic quasar continuum \citep[e.g][]{Francis1992,Suzuki2005,Durovcikova2020,Bosman2021_cont}.  The modelling of this intrinsic quasar property is a complex procedure, and can leave residual contamination in the transmitted flux fluctuations. This can be due to the diversity in quasar continua when employing statistical methods to reconstruct the continuum, or due to absorption that can remove a large fraction of the intrinsic flux.  The second of the two is especially of note for high-redshift quasars, where the larger IGM neutral fraction causes on average stronger absorption.  A third possibility is linked to the echelle spectrographs commonly used in high-resolution spectroscopy.  Observations made at higher echelle orders produce distortions in the spectrum over each order that are subsequently corrected for during flux calibration.  This procedure can leave residual fluctuations imprinted on the absorption features. As has already been pointed out elsewhere \citep[e.g.][]{McDonald2005,Irsic2017_XQ100,Karacayli2022}, any of the above mentioned sources of quasar continuum fluctuations could lead to enhanced power on larger scales, and this enhancement will be similar to the effect of patchy reionization.  With this specifically in mind, \citet{Karacayli2022} marginalized over the parameters of their quasar continuum model in order to mitigate any scale-dependence from variations in the continuum.  The preference for an enhancement in the large scale power we find in this work is therefore obtained  \emph{already assuming} a reasonable range of variation in the continuum.   The marginalization procedure used by \citet{Karacayli2022}, however, significantly increases the uncertainty on the power spectrum on large scales, and therefore effectively limits the increase in statistical power gained from combining different data sets.

There can also be other astrophysical effects that contribute to the shape of the large-scale power spectrum.  Outflows driven by active galactic nuclei can \emph{suppress} power on large scales, although this effect only becomes important toward lower redshifts, $z\lesssim 2.5$, where the volume filling factor of the outflows is larger and the \Lya forest is sensitive to higher density gas \citep{Viel2013_feedback,Chabanier2020}.   At higher redshifts, enhancements to the power spectrum on large scales are more likely to arise from high column density systems, such as damped \Lya absorbers and super Lyman limit systems with column densities $N_{\rm HI}\gtrsim 10^{19}\rm\,cm^{-2}$. These systems have large Lorentzian damping wings that correlate on scales of order of $\sim 1000\;\mathrm{s^{-1}\,km}$ or $\sim 10\,h^{-1}\mathrm{Mpc}$.  Measurements of the \Lya forest flux power spectrum therefore typically compile a catalogue of such systems, and the affected regions are subsequently masked \citep{Irsic2017_XQ100,Walther2019,Boera2019}, but at the cost of a loss of information and hence statistical power \citep{Palanque2020,Karacayli2022}.   However, due to incompleteness in the catalogues, particularly in low signal-to-noise data, residual effects from damping wings can remain.  This effect was studied in detail by \citet{Rogers2018} \citep[and see also][]{McDonald2005,FontRibera2012},  where it was found that damped systems introduce a large-scale enhancement in the power spectrum that is similar to that expected from patchy reionization. A more complex analysis would therefore seek to further marginalize over the parameters of a model for contamination by damped absorbers.  

The onset of inhomogeneous \HeII reionisation at redshifts $z \lesssim 4$ \citep[e.g.][]{Worseck2016} could also potentially impact on the \Lya forest power spectrum, although the effect on the 1D power spectrum at large scales is minimal \citep{McQuinn2011}.  The largest effect is instead expected at small scales due to increased line widths associated with \HeII photo-heating \citep{LaPlante2018,UptonSanderbeck2020}. \edit{Furthermore the bulk of the additional \HeII photo-heating should occur at lower redshifts than those we consider here, particularly if \HeII reionization does not fully complete until $z \sim 3$.}

\edit{Lastly, in this work we have assumed a fixed $\Lambda$CDM cosmology, but the enhancement of large-scale power by reionization also has important consequences for the inference of cosmological parameters from the \Lya forest power spectrum.  This will be relevant for parameters that derive most of the constraining power from large scales. For example, parameters that change the amplitude of matter fluctuations, such as $\sigma_{8}$ and the sum of the neutrino masses, $\Sigma_\nu$, could be biased high if the impact of reionization on large scales is not accounted for.   Higher precision measurements of the flux power spectrum at large scales could alleviate this tension, as the change in the shape of power spectrum will be different when varying either $\sigma_{8}$ or the reionization model.  However, such reasoning is more complicated for the value of the spectral index, $n_{\rm s}$, and the running of the spectral index, $\alpha_{\rm s}$.  A combination of a lower value for $n_{\rm s}$ and a lower mean \Lya forest transmission can mimic the shape of the reionization signal when averaged over too narrow a range of wavenumbers. Such an analysis therefore runs the risk of $n_{\rm s}$ being biased low, or instead trading a low $n_{\rm s}$ value for a non-zero running of the spectral index.  Further study of these effects will be required when inferring cosmological parameters from the \Lya forest power spectrum at $z>4$.}

\subsection{Implications for future constraints on inhomogeneous reionization}

\label{section:higher_pres} 
\begin{figure}
    \centering
        \includegraphics[trim=50 0 0 0, width=10cm ]{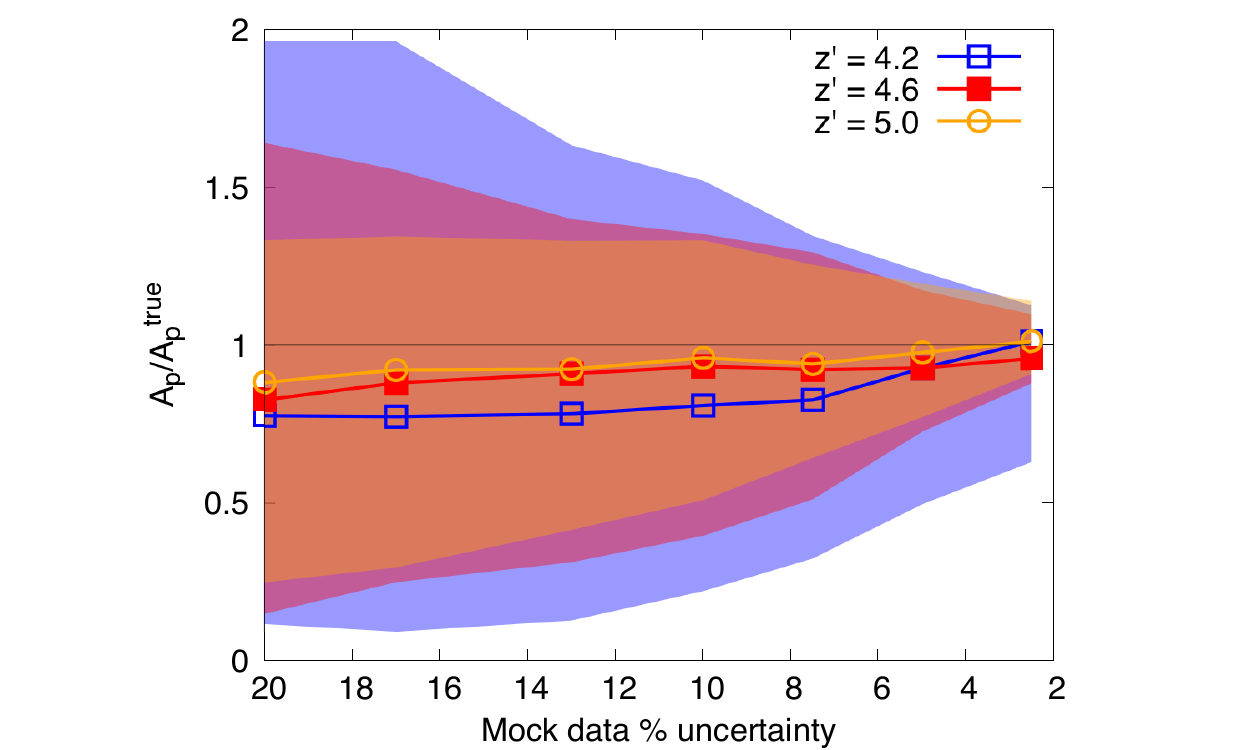}
    \caption{Median (solid curves) and 68 per cent confidence interval (shaded regions) for $A_{\rm p}/A_{\rm p}^{\rm true}$ obtained from $500$ mock realisations of the 1D \Lya forest power spectrum with varying relative uncertainties.  The results are shown separately for the three redshift bins considered. The solid grey line shows the perfect recovery case, where $A_{\rm p}/A_{\rm p}^{\rm true}=1$.}
    \label{fig:summary_Ap_and_zR}
\end{figure}

We now briefly turn to consider how future, high precision measurements of the \Lya forest power spectrum on large scales may be used detect the signature of inhomogeneous reionization.  We achieve this by constructing mock realisations of the power spectrum using the RT-late simulation described in M22.  We then vary the size of the relative uncertainty on the mock realisations, and apply our parameter inference framework to obtain $A_{\rm p}$.  This enables us to assess the precision that the true underlying value of $A_{\rm p}$ used in the simulation --  which we refer to as \Amatch{} -- is recovered from the mock data.  

In Fig.~\ref{fig:summary_Ap_and_zR}, we obtain a distribution for the recovered best fit \Ap{} parameter after performing our MCMC analysis assuming different relative uncertainties on the power spectrum measurements on 500 mock realisations.  We show the median (solid curves) and the 68 per cent confidence intervals (shaded regions) for $A_{\rm p}/A_{\rm p}^{\rm true}$ in three different redshift bins for different relative uncertainties on the power spectrum.  We start from a $20$ per cent relative uncertainty -- comparable to that in \citet{Karacayli2022} -- and extend this to higher precision cases that may be within reach of future surveys such as WEAVE-QSO and DESI.   Note the recovered 68 per cent scatter is smaller in the higher redshift bins because the large scale enhancement in the 1D power spectrum due to patchy reionization is more prominent approaching the end-point of reionization.

Fig.~\ref{fig:summary_Ap_and_zR} highlights how the precision of the \Amatch{} recovery significantly improves as the relative uncertainty on the mock data decreases.  For future, high-precision observations of the 1D \Lya power spectrum, this opens the possibility of using the power spectrum on large scales to unambiguously recover the signature of post-reionisation fluctuations in the IGM thermal state.  We expect that this will be most useful when combined with a joint analysis of the transmitted flux distribution \citep[e.g.][]{Bosman2022} to obtain self-consistent constraints on the timing of reionization.


\section{Conclusions} \label{section:conclusions}

Inhomogeneous reionization impacts on the shape of the 1D \Lya forest  power spectrum at redshifts $z\geq 4$.  This is because of coherent fluctuations in the ionized hydrogen fraction on large scales \citep[see also][]{Cen2009,Keating2018,Daloisio2019,Onorbe2019,Wu2019,Montero2020}, and spatial variations in the thermal broadening kernel and divergent peculiar velocities at small scales \citep{Wu2019,Onorbe2019,Molaro2022}.  

In this work, we have used the Sherwood-Relics  simulations \citep[see][for an overview]{Puchwein2022} to assess the possibility of detecting the relic signature of inhomogeneous reionization  at $4.0\leq z \leq 4.6$ using measurements of the 1D power spectrum at wavenumbers $\log_{10}(k/\rm km^{-1}\,s)\leq -2.2$ \citep{Karacayli2022}, corresponding to larger scales than those considered in recent analyses of the intergalactic medium (IGM) thermal state \citep[][although see also \citet{DAloisio2018fluc} for earlier work using data at higher redshift]{Boera2019, Molaro2022}.   We update the analysis framework introduced by \citet{Molaro2022} by introducing a neural-network-based interpolator in our Bayesian inference analysis.  This approach improves the control of interpolation uncertainties in our analysis (typically $< 1$ per cent), and is more computationally efficient than our previous linear interpolation technique.     We also use a more general, model independent approach for parameterising the power spectrum template for inhomogeneous reionization first presented by \citet{Molaro2022}.  Our main results are as follows:

\begin{itemize}

\item As already discussed by \citet{Molaro2022}, we find that if considering mock realisations of the \Lya forest power spectrum at wavenumbers $-2.2 \leq \log_{10} (k/{\rm km^{-1}\,s})\leq -0.7$ with 5 per cent relative uncertainties, patchy reionization introduces a modest $\sim 1\sigma$ shift between the true and recovered IGM parameters if (incorrectly) assuming a spatially uniform UV background.    However, when including larger scales with wavenumbers $\log_{10} (k/{\rm km^{-1}\,s}) \simeq -3$, the bias in the parameter recovery is much larger and is at the level of $\sim 6 \sigma$.  This demonstrates that IGM parameter recovery is significantly more sensitive to inhomogeneous reionization when including power spectrum data on large scales.\\

\item We perform a first analysis of the \Lya forest power spectrum measurements presented recently by \citet{Karacayli2022}, which are based on the combined observational data 
from the Keck Observatory Database of Ionized Absorption toward Quasars (KODIAQ) \citep{OMeara2017}, the Spectral Quasar Absorption Database (SQUAD) \citep{Murphy2019} and
the XQ-100 survey \citep{Lopez2016}.  These data cover four redshift bins at $z=4.0,4.2,4.4$, and $4.6$, and extend to large scales, $\log_{10} (k/\rm km^{-1}\,s) \simeq -3$ and have typical relative uncertainties of $10$--$20$ per cent in each wavenumber bin.   We find a preference at the $2.7\sigma$ level for an enhancement in the \citet{Karacayli2022} power spectrum at large scales relative to models that assume a spatially uniform background.  This additional power may be due to patchy reionization, although we caution that systematic effects (e.g. variations in the shape of the continuum placement, or the damping wings of high column density systems) could still contribute.\\

\item The precision of the \citet{Karacayli2022} power spectrum data at large scales is insufficient for recovering the signature of inhomogeneous reionization at high significance.  However, forthcoming surveys such as the Dark Energy Spectropscopic Instrument (DESI) survey \citep{VargasMagana2019} and William Herschel Telescope Enhanced Area Velocity Explorer QSO (WEAVE-QSO) survey \citep{Pieri2016} will also extend to large scales at $2\leq z\leq 4.5$, and will measure the 1D \Lya forest power spectrum to a precision of a few per cent.  Any enhancement in the power spectrum on large scales should then be recoverable.   We anticipate that combining the 1D power spectrum on large scales in a joint analysis with the transmitted flux distribution \citep[e.g.][]{Bosman2022} should provide a powerful constraint on the timing of reionization.

\end{itemize}

\section*{Acknowledgements}
\edit{The authors would like to thank the anonymous referee for very helpful comments.} We thank Steven Bamford for very useful discussions. The simulations used in this work were performed using the Joliot Curie supercomputer at the \edit{Trés} Grand Centre de Calcul (TGCC) and the Cambridge Service for Data Driven Discovery (CSD3), part of which is operated by the University of Cambridge Research Computing on behalf of the STFC DiRAC HPC Facility (www.dirac.ac.uk).  We acknowledge the Partnership for Advanced Computing in Europe (PRACE) for awarding us time on Joliot Curie in the 16th call. The DiRAC component of CSD3 was funded by BEIS capital funding via STFC capital grants ST/P002307/1 and ST/R002452/1 and STFC operations grant ST/R00689X/1.  This work also used the DiRAC@Durham facility managed by the Institute for Computational Cosmology on behalf of the STFC DiRAC HPC Facility. The equipment was funded by BEIS capital funding via STFC capital grants ST/P002293/1 and ST/R002371/1, Durham University and STFC operations grant ST/R000832/1. DiRAC is part of the National e-Infrastructure.   MM and JSB are supported by STFC consolidated grant ST/T000171/1. VI is supported by the Kavli foundation.  MGH acknowledges support from the UKRI STFC (grant numbers ST/N000927/1 and ST/S000623/1).  LCK was supported by the European Union’s Horizon 2020 research 
and innovation programme under the Marie Skłodowska-Curie grant agreement 
No. 885990. We thank Volker Springel for making \textsc{P-GADGET-3} available. We also thank Dominique Aubert for sharing the ATON code. We acknowledge use of the keras2c library in this work \citep{Keras2c_ref}. \edit{For the purpose of open access, the author has applied a Creative Commons Attribution (CC BY) licence to any Author Accepted Manuscript version arising from this submission.}


\section*{Data Availability}
All data and analysis code used in this work are available from the first author on reasonable request.  



\bibliographystyle{mnras}
\bibliography{references_this}

\begin{thebibliography}{}
\makeatletter
\relax
\def\mn@urlcharsother{\let\do\@makeother \do\$\do\&\do\#\do\^\do\_\do\%\do\~}
\def\mn@doi{\begingroup\mn@urlcharsother \@ifnextchar [ {\mn@doi@}
  {\mn@doi@[]}}
\def\mn@doi@[#1]#2{\def\@tempa{#1}\ifx\@tempa\@empty \href
  {http://dx.doi.org/#2} {doi:#2}\else \href {http://dx.doi.org/#2} {#1}\fi
  \endgroup}
\def\mn@eprint#1#2{\mn@eprint@#1:#2::\@nil}
\def\mn@eprint@arXiv#1{\href {http://arxiv.org/abs/#1} {{\tt arXiv:#1}}}
\def\mn@eprint@dblp#1{\href {http://dblp.uni-trier.de/rec/bibtex/#1.xml}
  {dblp:#1}}
\def\mn@eprint@#1:#2:#3:#4\@nil{\def\@tempa {#1}\def\@tempb {#2}\def\@tempc
  {#3}\ifx \@tempc \@empty \let \@tempc \@tempb \let \@tempb \@tempa \fi \ifx
  \@tempb \@empty \def\@tempb {arXiv}\fi \@ifundefined
  {mn@eprint@\@tempb}{\@tempb:\@tempc}{\expandafter \expandafter \csname
  mn@eprint@\@tempb\endcsname \expandafter{\@tempc}}}

\bibitem[\protect\citeauthoryear{{Becker}, {Bolton}, {Madau}, {Pettini},
  {Ryan-Weber}  \& {Venemans}}{{Becker} et~al.}{2015}]{Becker2015}
{Becker} G.~D.,  {Bolton} J.~S.,  {Madau} P.,  {Pettini} M.,  {Ryan-Weber}
  E.~V.,   {Venemans} B.~P.,  2015, \mn@doi [\mnras] {10.1093/mnras/stu2646},
  \href {https://ui.adsabs.harvard.edu/abs/2015MNRAS.447.3402B} {447, 3402}

\bibitem[\protect\citeauthoryear{{Bird}, {Rogers}, {Peiris}, {Verde},
  {Font-Ribera}  \& {Pontzen}}{{Bird} et~al.}{2019}]{Bird2019}
{Bird} S.,  {Rogers} K.~K.,  {Peiris} H.~V.,  {Verde} L.,  {Font-Ribera} A.,
  {Pontzen} A.,  2019, \mn@doi [\jcap] {10.1088/1475-7516/2019/02/050}, \href
  {https://ui.adsabs.harvard.edu/abs/2019JCAP...02..050B} {2019, 050}

\bibitem[\protect\citeauthoryear{{Boera}, {Murphy}, {Becker}  \&
  {Bolton}}{{Boera} et~al.}{2014}]{Boera2014}
{Boera} E.,  {Murphy} M.~T.,  {Becker} G.~D.,   {Bolton} J.~S.,  2014, \mn@doi
  [\mnras] {10.1093/mnras/stu660}, \href
  {https://ui.adsabs.harvard.edu/abs/2014MNRAS.441.1916B} {441, 1916}

\bibitem[\protect\citeauthoryear{{Boera}, {Becker}, {Bolton}  \&
  {Nasir}}{{Boera} et~al.}{2019}]{Boera2019}
{Boera} E.,  {Becker} G.~D.,  {Bolton} J.~S.,   {Nasir} F.,  2019, \mn@doi
  [\apj] {10.3847/1538-4357/aafee4}, \href
  {https://ui.adsabs.harvard.edu/abs/2019ApJ...872..101B} {872, 101}

\bibitem[\protect\citeauthoryear{{Bolton}, {Haehnelt}, {Viel}  \&
  {Springel}}{{Bolton} et~al.}{2005}]{Bolton2005}
{Bolton} J.~S.,  {Haehnelt} M.~G.,  {Viel} M.,   {Springel} V.,  2005, \mn@doi
  [\mnras] {10.1111/j.1365-2966.2005.08704.x}, \href
  {https://ui.adsabs.harvard.edu/abs/2005MNRAS.357.1178B} {357, 1178}

\bibitem[\protect\citeauthoryear{{Bosman}, {{\v{D}}urov{\v{c}}{\'\i}kov{\'a}},
  {Davies}  \& {Eilers}}{{Bosman} et~al.}{2021}]{Bosman2021_cont}
{Bosman} S. E.~I.,  {{\v{D}}urov{\v{c}}{\'\i}kov{\'a}} D.,  {Davies} F.~B.,
  {Eilers} A.-C.,  2021, \mn@doi [\mnras] {10.1093/mnras/stab572}, \href
  {https://ui.adsabs.harvard.edu/abs/2021MNRAS.503.2077B} {503, 2077}

\bibitem[\protect\citeauthoryear{{Bosman} et~al.,}{{Bosman}
  et~al.}{2022}]{Bosman2022}
{Bosman} S. E.~I.,  et~al., 2022, \mn@doi [\mnras] {10.1093/mnras/stac1046},
  \href {https://ui.adsabs.harvard.edu/abs/2022MNRAS.514...55B} {514, 55}

\bibitem[\protect\citeauthoryear{Cen, McDonald, Trac  \& Loeb}{Cen
  et~al.}{2009}]{Cen2009}
Cen R.,  McDonald P.,  Trac H.,   Loeb A.,  2009, \mn@doi [The Astrophysical
  Journal] {10.1088/0004-637x/706/1/l164}, 706, L164

\bibitem[\protect\citeauthoryear{{Chabanier} et~al.,}{{Chabanier}
  et~al.}{2019}]{Chabanier2019}
{Chabanier} S.,  et~al., 2019, \mn@doi [\jcap] {10.1088/1475-7516/2019/07/017},
  \href {https://ui.adsabs.harvard.edu/abs/2019JCAP...07..017C} {2019, 017}

\bibitem[\protect\citeauthoryear{{Chabanier}, {Bournaud}, {Dubois},
  {Palanque-Delabrouille}, {Y{\`e}che}, {Armengaud}, {Peirani}  \&
  {Beckmann}}{{Chabanier} et~al.}{2020}]{Chabanier2020}
{Chabanier} S.,  {Bournaud} F.,  {Dubois} Y.,  {Palanque-Delabrouille} N.,
  {Y{\`e}che} C.,  {Armengaud} E.,  {Peirani} S.,   {Beckmann} R.,  2020,
  \mn@doi [\mnras] {10.1093/mnras/staa1242}, \href
  {https://ui.adsabs.harvard.edu/abs/2020MNRAS.495.1825C} {495, 1825}

\bibitem[\protect\citeauthoryear{Conlin, Erickson, Abbate  \& Kolemen}{Conlin
  et~al.}{2021}]{Keras2c_ref}
Conlin R.,  Erickson K.,  Abbate J.,   Kolemen E.,  2021, \mn@doi [Engineering
  Applications of Artificial Intelligence] {10.1016/j.engappai.2021.104182},
  100

\bibitem[\protect\citeauthoryear{{Croft}, {Weinberg}, {Bolte}, {Burles},
  {Hernquist}, {Katz}, {Kirkman}  \& {Tytler}}{{Croft}
  et~al.}{2002}]{Croft2002}
{Croft} R. A.~C.,  {Weinberg} D.~H.,  {Bolte} M.,  {Burles} S.,  {Hernquist}
  L.,  {Katz} N.,  {Kirkman} D.,   {Tytler} D.,  2002, \mn@doi [\apj]
  {10.1086/344099}, \href
  {https://ui.adsabs.harvard.edu/abs/2002ApJ...581...20C} {581, 20}

\bibitem[\protect\citeauthoryear{{D'Aloisio}, {McQuinn}, {Davies}  \&
  {Furlanetto}}{{D'Aloisio} et~al.}{2018}]{DAloisio2018fluc}
{D'Aloisio} A.,  {McQuinn} M.,  {Davies} F.~B.,   {Furlanetto} S.~R.,  2018,
  \mn@doi [\mnras] {10.1093/mnras/stx2341}, \href
  {https://ui.adsabs.harvard.edu/abs/2018MNRAS.473..560D} {473, 560}

\bibitem[\protect\citeauthoryear{D'Aloisio, McQuinn, Maupin, Davies, Trac,
  Fuller  \& Sanderbeck}{D'Aloisio et~al.}{2019}]{Daloisio2019}
D'Aloisio A.,  McQuinn M.,  Maupin O.,  Davies F.~B.,  Trac H.,  Fuller S.,
  Sanderbeck P. R.~U.,  2019, \mn@doi [The Astrophysical Journal]
  {10.3847/1538-4357/ab0d83}, 874, 154

\bibitem[\protect\citeauthoryear{{Eilers}, {Davies}  \& {Hennawi}}{{Eilers}
  et~al.}{2018}]{Eilers2018}
{Eilers} A.-C.,  {Davies} F.~B.,   {Hennawi} J.~F.,  2018, \mn@doi [\apj]
  {10.3847/1538-4357/aad4fd}, \href
  {https://ui.adsabs.harvard.edu/abs/2018ApJ...864...53E} {864, 53}

\bibitem[\protect\citeauthoryear{{Fan}, {Carilli}  \& {Keating}}{{Fan}
  et~al.}{2006}]{Fan2006}
{Fan} X.,  {Carilli} C.~L.,   {Keating} B.,  2006, \mn@doi [\araa]
  {10.1146/annurev.astro.44.051905.092514}, \href
  {http://adsabs.harvard.edu/abs/2006ARA%26A..44..415F} {44, 415}

\bibitem[\protect\citeauthoryear{{Fernandez}, {Ho}  \& {Bird}}{{Fernandez}
  et~al.}{2022}]{Fernandez2022}
{Fernandez} M.~A.,  {Ho} M.-F.,   {Bird} S.,  2022, arXiv e-prints, \href
  {https://ui.adsabs.harvard.edu/abs/2022arXiv220706445F} {p. arXiv:2207.06445}

\bibitem[\protect\citeauthoryear{{Font-Ribera} \&
  {Miralda-Escud{\'e}}}{{Font-Ribera} \&
  {Miralda-Escud{\'e}}}{2012}]{FontRibera2012}
{Font-Ribera} A.,  {Miralda-Escud{\'e}} J.,  2012, \mn@doi [\jcap]
  {10.1088/1475-7516/2012/07/028}, \href
  {https://ui.adsabs.harvard.edu/abs/2012JCAP...07..028F} {2012, 028}

\bibitem[\protect\citeauthoryear{{Francis}, {Hewett}, {Foltz}  \&
  {Chaffee}}{{Francis} et~al.}{1992}]{Francis1992}
{Francis} P.~J.,  {Hewett} P.~C.,  {Foltz} C.~B.,   {Chaffee} F.~H.,  1992,
  \mn@doi [\apj] {10.1086/171870}, \href
  {https://ui.adsabs.harvard.edu/abs/1992ApJ...398..476F} {398, 476}

\bibitem[\protect\citeauthoryear{Fukushima}{Fukushima}{1975}]{fukushima1975cognitron}
Fukushima K.,  1975, Biological cybernetics, 20, 121

\bibitem[\protect\citeauthoryear{{Gaikwad} et~al.,}{{Gaikwad}
  et~al.}{2020}]{Gaikwad2020}
{Gaikwad} P.,  et~al., 2020, \mn@doi [\mnras] {10.1093/mnras/staa907}, \href
  {https://ui.adsabs.harvard.edu/abs/2020MNRAS.494.5091G} {494, 5091}

\bibitem[\protect\citeauthoryear{{Gaikwad}, {Srianand}, {Haehnelt}  \&
  {Choudhury}}{{Gaikwad} et~al.}{2021}]{Gaikwad2021}
{Gaikwad} P.,  {Srianand} R.,  {Haehnelt} M.~G.,   {Choudhury} T.~R.,  2021,
  \mn@doi [\mnras] {10.1093/mnras/stab2017}, \href
  {https://ui.adsabs.harvard.edu/abs/2021MNRAS.506.4389G} {506, 4389}

\bibitem[\protect\citeauthoryear{{Garzilli}, {Magalich}, {Theuns}, {Frenk},
  {Weniger}, {Ruchayskiy}  \& {Boyarsky}}{{Garzilli}
  et~al.}{2019}]{Garzilli2019}
{Garzilli} A.,  {Magalich} A.,  {Theuns} T.,  {Frenk} C.~S.,  {Weniger} C.,
  {Ruchayskiy} O.,   {Boyarsky} A.,  2019, \mn@doi [\mnras]
  {10.1093/mnras/stz2188}, \href
  {https://ui.adsabs.harvard.edu/abs/2019MNRAS.489.3456G} {489, 3456}

\bibitem[\protect\citeauthoryear{{Gurvich}, {Burkhart}  \& {Bird}}{{Gurvich}
  et~al.}{2017}]{Gurvich2017}
{Gurvich} A.,  {Burkhart} B.,   {Bird} S.,  2017, \mn@doi [\apj]
  {10.3847/1538-4357/835/2/175}, \href
  {https://ui.adsabs.harvard.edu/abs/2017ApJ...835..175G} {835, 175}

\bibitem[\protect\citeauthoryear{{Hiss}, {Walther}, {Hennawi}, {O{\~n}orbe},
  {O'Meara}, {Rorai}  \& {Luki{\'c}}}{{Hiss} et~al.}{2018}]{Hiss2018}
{Hiss} H.,  {Walther} M.,  {Hennawi} J.~F.,  {O{\~n}orbe} J.,  {O'Meara} J.~M.,
   {Rorai} A.,   {Luki{\'c}} Z.,  2018, \mn@doi [\apj]
  {10.3847/1538-4357/aada86}, \href
  {https://ui.adsabs.harvard.edu/abs/2018ApJ...865...42H} {865, 42}

\bibitem[\protect\citeauthoryear{Hornik, Stinchcombe  \& White}{Hornik
  et~al.}{1989}]{hornik1989multilayer}
Hornik K.,  Stinchcombe M.,   White H.,  1989, Neural networks, 2, 359

\bibitem[\protect\citeauthoryear{{Hsyu}, {Cooke}, {Prochaska}  \&
  {Bolte}}{{Hsyu} et~al.}{2020}]{Hsyu2020}
{Hsyu} T.,  {Cooke} R.~J.,  {Prochaska} J.~X.,   {Bolte} M.,  2020, \mn@doi
  [\apj] {10.3847/1538-4357/ab91af}, 896, 77

\bibitem[\protect\citeauthoryear{{Hui} \& {Gnedin}}{{Hui} \&
  {Gnedin}}{1997}]{Hui1997}
{Hui} L.,  {Gnedin} N.~Y.,  1997, \mn@doi [\mnras] {10.1093/mnras/292.1.27},
  \href {https://ui.adsabs.harvard.edu/abs/1997MNRAS.292...27H} {292, 27}

\bibitem[\protect\citeauthoryear{{Hui} \& {Haiman}}{{Hui} \&
  {Haiman}}{2003}]{HuiHaiman2003}
{Hui} L.,  {Haiman} Z.,  2003, \mn@doi [\apj] {10.1086/377229}, \href
  {https://ui.adsabs.harvard.edu/abs/2003ApJ...596....9H} {596, 9}

\bibitem[\protect\citeauthoryear{{Ir{\v{s}}i{\v{c}}}
  et~al.,}{{Ir{\v{s}}i{\v{c}}} et~al.}{2017a}]{Irsic2017}
{Ir{\v{s}}i{\v{c}}} V.,  et~al., 2017a, \mn@doi [\prd]
  {10.1103/PhysRevD.96.023522}, \href
  {https://ui.adsabs.harvard.edu/abs/2017PhRvD..96b3522I} {96, 023522}

\bibitem[\protect\citeauthoryear{{Ir{\v{s}}i{\v{c}}}
  et~al.,}{{Ir{\v{s}}i{\v{c}}} et~al.}{2017b}]{Irsic2017_XQ100}
{Ir{\v{s}}i{\v{c}}} V.,  et~al., 2017b, \mn@doi [\mnras]
  {10.1093/mnras/stw3372}, \href
  {https://ui.adsabs.harvard.edu/abs/2017MNRAS.466.4332I} {466, 4332}

\bibitem[\protect\citeauthoryear{{Kara{\c{c}}ayl{\i}}, {Font-Ribera}  \&
  {Padmanabhan}}{{Kara{\c{c}}ayl{\i}} et~al.}{2020}]{Karacayli2020}
{Kara{\c{c}}ayl{\i}} N.~G.,  {Font-Ribera} A.,   {Padmanabhan} N.,  2020,
  \mn@doi [\mnras] {10.1093/mnras/staa2331}, \href
  {https://ui.adsabs.harvard.edu/abs/2020MNRAS.497.4742K} {497, 4742}

\bibitem[\protect\citeauthoryear{{Kara{\c{c}}ayl{\i}}
  et~al.,}{{Kara{\c{c}}ayl{\i}} et~al.}{2022}]{Karacayli2022}
{Kara{\c{c}}ayl{\i}} N.~G.,  et~al., 2022, \mn@doi [\mnras]
  {10.1093/mnras/stab3201}, \href
  {https://ui.adsabs.harvard.edu/abs/2022MNRAS.509.2842K} {509, 2842}

\bibitem[\protect\citeauthoryear{{Keating}, {Puchwein}  \&
  {Haehnelt}}{{Keating} et~al.}{2018}]{Keating2018}
{Keating} L.~C.,  {Puchwein} E.,   {Haehnelt} M.~G.,  2018, \mn@doi [\mnras]
  {10.1093/mnras/sty968}, \href
  {https://ui.adsabs.harvard.edu/abs/2018MNRAS.477.5501K} {477, 5501}

\bibitem[\protect\citeauthoryear{{La Plante}, {Trac}, {Croft}  \& {Cen}}{{La
  Plante} et~al.}{2018}]{LaPlante2018}
{La Plante} P.,  {Trac} H.,  {Croft} R.,   {Cen} R.,  2018, \mn@doi [\apj]
  {10.3847/1538-4357/aae693}, \href
  {https://ui.adsabs.harvard.edu/abs/2018ApJ...868..106L} {868, 106}

\bibitem[\protect\citeauthoryear{LeCun, Bottou, Orr  \& M{\"u}ller}{LeCun
  et~al.}{2012}]{lecun2012efficient}
LeCun Y.~A.,  Bottou L.,  Orr G.~B.,   M{\"u}ller K.-R.,  2012, in , Neural
  networks: Tricks of the trade.
Springer, pp 9--48

\bibitem[\protect\citeauthoryear{{L{\'o}pez} et~al.,}{{L{\'o}pez}
  et~al.}{2016}]{Lopez2016}
{L{\'o}pez} S.,  et~al., 2016, \mn@doi [\aap] {10.1051/0004-6361/201628161},
  \href {https://ui.adsabs.harvard.edu/abs/2016A&A...594A..91L} {594, A91}

\bibitem[\protect\citeauthoryear{{McDonald}, {Miralda-Escud{\'e}}, {Rauch},
  {Sargent}, {Barlow}, {Cen}  \& {Ostriker}}{{McDonald}
  et~al.}{2000}]{McDonald2000}
{McDonald} P.,  {Miralda-Escud{\'e}} J.,  {Rauch} M.,  {Sargent} W. L.~W.,
  {Barlow} T.~A.,  {Cen} R.,   {Ostriker} J.~P.,  2000, \mn@doi [\apj]
  {10.1086/317079}, \href
  {https://ui.adsabs.harvard.edu/abs/2000ApJ...543....1M} {543, 1}

\bibitem[\protect\citeauthoryear{{McDonald}, {Seljak}, {Cen}, {Bode}  \&
  {Ostriker}}{{McDonald} et~al.}{2005}]{McDonald2005}
{McDonald} P.,  {Seljak} U.,  {Cen} R.,  {Bode} P.,   {Ostriker} J.~P.,  2005,
  \mn@doi [\mnras] {10.1111/j.1365-2966.2005.09141.x}, \href
  {https://ui.adsabs.harvard.edu/abs/2005MNRAS.360.1471M} {360, 1471}

\bibitem[\protect\citeauthoryear{{McGreer}, {Mesinger}  \&
  {D'Odorico}}{{McGreer} et~al.}{2015}]{McGreer2015}
{McGreer} I.~D.,  {Mesinger} A.,   {D'Odorico} V.,  2015, \mn@doi [\mnras]
  {10.1093/mnras/stu2449}, \href
  {https://ui.adsabs.harvard.edu/abs/2015MNRAS.447..499M} {447, 499}

\bibitem[\protect\citeauthoryear{{McQuinn}}{{McQuinn}}{2016}]{McQuinn2016}
{McQuinn} M.,  2016, \mn@doi [\araa] {10.1146/annurev-astro-082214-122355},
  \href {https://ui.adsabs.harvard.edu/abs/2016ARA&A..54..313M} {54, 313}

\bibitem[\protect\citeauthoryear{{McQuinn}, {Hernquist}, {Lidz}  \&
  {Zaldarriaga}}{{McQuinn} et~al.}{2011}]{McQuinn2011}
{McQuinn} M.,  {Hernquist} L.,  {Lidz} A.,   {Zaldarriaga} M.,  2011, \mn@doi
  [\mnras] {10.1111/j.1365-2966.2011.18788.x}, \href
  {https://ui.adsabs.harvard.edu/abs/2011MNRAS.415..977M} {415, 977}

\bibitem[\protect\citeauthoryear{{Mishra} \& {Gnedin}}{{Mishra} \&
  {Gnedin}}{2022}]{Mishra2022}
{Mishra} N.,  {Gnedin} N.~Y.,  2022, \mn@doi [\apj] {10.3847/1538-4357/ac5a50},
  \href {https://ui.adsabs.harvard.edu/abs/2022ApJ...928..174M} {928, 174}

\bibitem[\protect\citeauthoryear{{Molaro} et~al.,}{{Molaro}
  et~al.}{2022}]{Molaro2022}
{Molaro} M.,  et~al., 2022, \mn@doi [\mnras] {10.1093/mnras/stab3416}, \href
  {https://ui.adsabs.harvard.edu/abs/2022MNRAS.509.6119M} {509, 6119}

\bibitem[\protect\citeauthoryear{{Montero-Camacho} \& {Mao}}{{Montero-Camacho}
  \& {Mao}}{2020}]{Montero2020}
{Montero-Camacho} P.,  {Mao} Y.,  2020, \mn@doi [\mnras]
  {10.1093/mnras/staa2918}, \href
  {https://ui.adsabs.harvard.edu/abs/2020MNRAS.499.1640M} {499, 1640}

\bibitem[\protect\citeauthoryear{{Murphy}, {Kacprzak}, {Savorgnan}  \&
  {Carswell}}{{Murphy} et~al.}{2019}]{Murphy2019}
{Murphy} M.~T.,  {Kacprzak} G.~G.,  {Savorgnan} G. A.~D.,   {Carswell} R.~F.,
  2019, \mn@doi [\mnras] {10.1093/mnras/sty2834}, \href
  {https://ui.adsabs.harvard.edu/abs/2019MNRAS.482.3458M} {482, 3458}

\bibitem[\protect\citeauthoryear{{Nasir}, {Bolton}  \& {Becker}}{{Nasir}
  et~al.}{2016}]{Nasir2016}
{Nasir} F.,  {Bolton} J.~S.,   {Becker} G.~D.,  2016, \mn@doi [\mnras]
  {10.1093/mnras/stw2147}, \href
  {https://ui.adsabs.harvard.edu/abs/2016MNRAS.463.2335N} {463, 2335}

\bibitem[\protect\citeauthoryear{{O{\~n}orbe}, {Davies}, {Luki{\'c}}, {},
  {Hennawi}  \& {Sorini}}{{O{\~n}orbe} et~al.}{2019}]{Onorbe2019}
{O{\~n}orbe} J.,  {Davies} F.~B.,  {Luki{\'c}} {} Z.,  {Hennawi} J.~F.,
  {Sorini} D.,  2019, \mn@doi [\mnras] {10.1093/mnras/stz984}, 486, 4075

\bibitem[\protect\citeauthoryear{{O'Meara}, {Lehner}, {Howk}, {Prochaska},
  {Fox}, {Peeples}, {Tumlinson}  \& {O'Shea}}{{O'Meara}
  et~al.}{2017}]{OMeara2017}
{O'Meara} J.~M.,  {Lehner} N.,  {Howk} J.~C.,  {Prochaska} J.~X.,  {Fox} A.~J.,
   {Peeples} M.~S.,  {Tumlinson} J.,   {O'Shea} B.~W.,  2017, \mn@doi [\aj]
  {10.3847/1538-3881/aa82b8}, \href
  {https://ui.adsabs.harvard.edu/abs/2017AJ....154..114O} {154, 114}

\bibitem[\protect\citeauthoryear{{Palanque-Delabrouille}
  et~al.,}{{Palanque-Delabrouille} et~al.}{2013}]{Palanque2013}
{Palanque-Delabrouille} N.,  et~al., 2013, \mn@doi [\aap]
  {10.1051/0004-6361/201322130}, \href
  {https://ui.adsabs.harvard.edu/abs/2013A&A...559A..85P} {559, A85}

\bibitem[\protect\citeauthoryear{{Palanque-Delabrouille}, {Y{\`e}che},
  {Sch{\"o}neberg}, {Lesgourgues}, {Walther}, {Chabanier}  \&
  {Armengaud}}{{Palanque-Delabrouille} et~al.}{2020}]{Palanque2020}
{Palanque-Delabrouille} N.,  {Y{\`e}che} C.,  {Sch{\"o}neberg} N.,
  {Lesgourgues} J.,  {Walther} M.,  {Chabanier} S.,   {Armengaud} E.,  2020,
  \mn@doi [\jcap] {10.1088/1475-7516/2020/04/038}, \href
  {https://ui.adsabs.harvard.edu/abs/2020JCAP...04..038P} {2020, 038}

\bibitem[\protect\citeauthoryear{Pan \& Yang}{Pan \&
  Yang}{2009}]{pan2009survey}
Pan S.~J.,  Yang Q.,  2009, IEEE Transactions on knowledge and data
  engineering, 22, 1345

\bibitem[\protect\citeauthoryear{{Pedersen}, {Font-Ribera}, {Rogers},
  {McDonald}, {Peiris}, {Pontzen}  \& {Slosar}}{{Pedersen}
  et~al.}{2021}]{Pedersen2021}
{Pedersen} C.,  {Font-Ribera} A.,  {Rogers} K.~K.,  {McDonald} P.,  {Peiris}
  H.~V.,  {Pontzen} A.,   {Slosar} A.,  2021, \mn@doi [\jcap]
  {10.1088/1475-7516/2021/05/033}, \href
  {https://ui.adsabs.harvard.edu/abs/2021JCAP...05..033P} {2021, 033}

\bibitem[\protect\citeauthoryear{{Pedersen}, {Font-Ribera}  \&
  {Gnedin}}{{Pedersen} et~al.}{2022}]{Pedersen2022}
{Pedersen} C.,  {Font-Ribera} A.,   {Gnedin} N.~Y.,  2022, arXiv e-prints,
  \href {https://ui.adsabs.harvard.edu/abs/2022arXiv220909895P} {p.
  arXiv:2209.09895}

\bibitem[\protect\citeauthoryear{{Pieri} et~al.,}{{Pieri}
  et~al.}{2016}]{Pieri2016}
{Pieri} M.~M.,  et~al., 2016, in {Reyl{\'e}} C.,  {Richard} J.,  {Cambr{\'e}sy}
  L.,  {Deleuil} M.,  {P{\'e}contal} E.,  {Tresse} L.,   {Vauglin} I.,  eds,
  SF2A-2016: Proceedings of the Annual meeting of the French Society of
  Astronomy and Astrophysics. pp 259--266 (\mn@eprint {arXiv} {1611.09388})

\bibitem[\protect\citeauthoryear{{Planck Collaboration} et~al.,}{{Planck
  Collaboration} et~al.}{2014}]{planck2014}
{Planck Collaboration} et~al., 2014, \mn@doi [A&A]
  {10.1051/0004-6361/201321591}, 571, A16

\bibitem[\protect\citeauthoryear{{Puchwein}, {Haardt}, {Haehnelt}  \&
  {Madau}}{{Puchwein} et~al.}{2019}]{Puchwein2019}
{Puchwein} E.,  {Haardt} F.,  {Haehnelt} M.~G.,   {Madau} P.,  2019, \mn@doi
  [\mnras] {10.1093/mnras/stz222}, \href
  {https://ui.adsabs.harvard.edu/abs/2019MNRAS.485...47P} {485, 47}

\bibitem[\protect\citeauthoryear{{Puchwein} et~al.,}{{Puchwein}
  et~al.}{2022}]{Puchwein2022}
{Puchwein} E.,  et~al., 2022, arXiv e-prints, \href
  {https://ui.adsabs.harvard.edu/abs/2022arXiv220713098P} {p. arXiv:2207.13098}

\bibitem[\protect\citeauthoryear{{Rauch}}{{Rauch}}{1998}]{Rauch1998}
{Rauch} M.,  1998, \mn@doi [\araa] {10.1146/annurev.astro.36.1.267}, \href
  {https://ui.adsabs.harvard.edu/abs/1998ARA&A..36..267R} {36, 267}

\bibitem[\protect\citeauthoryear{{Rogers} \& {Peiris}}{{Rogers} \&
  {Peiris}}{2021}]{Rogers2021}
{Rogers} K.~K.,  {Peiris} H.~V.,  2021, \mn@doi [\prl]
  {10.1103/PhysRevLett.126.071302}, \href
  {https://ui.adsabs.harvard.edu/abs/2021PhRvL.126g1302R} {126, 071302}

\bibitem[\protect\citeauthoryear{{Rogers}, {Bird}, {Peiris}, {Pontzen},
  {Font-Ribera}  \& {Leistedt}}{{Rogers} et~al.}{2018}]{Rogers2018}
{Rogers} K.~K.,  {Bird} S.,  {Peiris} H.~V.,  {Pontzen} A.,  {Font-Ribera} A.,
   {Leistedt} B.,  2018, \mn@doi [\mnras] {10.1093/mnras/stx2942}, \href
  {https://ui.adsabs.harvard.edu/abs/2018MNRAS.474.3032R} {474, 3032}

\bibitem[\protect\citeauthoryear{{Schaye}, {Theuns}, {Rauch}, {Efstathiou}  \&
  {Sargent}}{{Schaye} et~al.}{2000}]{Schaye2000}
{Schaye} J.,  {Theuns} T.,  {Rauch} M.,  {Efstathiou} G.,   {Sargent} W. L.~W.,
   2000, \mn@doi [\mnras] {10.1046/j.1365-8711.2000.03815.x}, \href
  {https://ui.adsabs.harvard.edu/abs/2000MNRAS.318..817S} {318, 817}

\bibitem[\protect\citeauthoryear{{Seljak}, {Makarov}, {McDonald}  \&
  {Trac}}{{Seljak} et~al.}{2006}]{Seljak2006}
{Seljak} U.,  {Makarov} A.,  {McDonald} P.,   {Trac} H.,  2006, \mn@doi [\prl]
  {10.1103/PhysRevLett.97.191303}, \href
  {https://ui.adsabs.harvard.edu/abs/2006PhRvL..97s1303S} {97, 191303}

\bibitem[\protect\citeauthoryear{Stone}{Stone}{1974}]{stone1974cross}
Stone M.,  1974, Journal of the royal statistical society: Series B
  (Methodological), 36, 111

\bibitem[\protect\citeauthoryear{{Suzuki}, {Tytler}, {Kirkman}, {O'Meara}  \&
  {Lubin}}{{Suzuki} et~al.}{2005}]{Suzuki2005}
{Suzuki} N.,  {Tytler} D.,  {Kirkman} D.,  {O'Meara} J.~M.,   {Lubin} D.,
  2005, \mn@doi [\apj] {10.1086/426062}, \href
  {https://ui.adsabs.harvard.edu/abs/2005ApJ...618..592S} {618, 592}

\bibitem[\protect\citeauthoryear{{Theuns}, {Schaye}, {Zaroubi}, {Kim},
  {Tzanavaris}  \& {Carswell}}{{Theuns} et~al.}{2002a}]{Theuns2002}
{Theuns} T.,  {Schaye} J.,  {Zaroubi} S.,  {Kim} T.-S.,  {Tzanavaris} P.,
  {Carswell} B.,  2002a, \mn@doi [\apjl] {10.1086/339998}, \href
  {https://ui.adsabs.harvard.edu/abs/2002ApJ...567L.103T} {567, L103}

\bibitem[\protect\citeauthoryear{{Theuns}, {Viel}, {Kay}, {Schaye}, {Carswell}
  \& {Tzanavaris}}{{Theuns} et~al.}{2002b}]{Theuns2002_winds}
{Theuns} T.,  {Viel} M.,  {Kay} S.,  {Schaye} J.,  {Carswell} R.~F.,
  {Tzanavaris} P.,  2002b, \mn@doi [\apjl] {10.1086/344521}, \href
  {https://ui.adsabs.harvard.edu/abs/2002ApJ...578L...5T} {578, L5}

\bibitem[\protect\citeauthoryear{{Upton Sanderbeck} \& {Bird}}{{Upton
  Sanderbeck} \& {Bird}}{2020}]{UptonSanderbeck2020}
{Upton Sanderbeck} P.,  {Bird} S.,  2020, \mn@doi [\mnras]
  {10.1093/mnras/staa1850}, \href
  {https://ui.adsabs.harvard.edu/abs/2020MNRAS.496.4372U} {496, 4372}

\bibitem[\protect\citeauthoryear{{Vargas-Magana}, {Brooks}, {Levi}  \&
  {Tarle}}{{Vargas-Magana} et~al.}{2019}]{VargasMagana2019}
{Vargas-Magana} M.,  {Brooks} D.~D.,  {Levi} M.~M.,   {Tarle} G.~G.,  2019,
  arXiv e-prints, \href {https://ui.adsabs.harvard.edu/abs/2019arXiv190101581V}
  {p. arXiv:1901.01581}

\bibitem[\protect\citeauthoryear{{Viel} \& {Haehnelt}}{{Viel} \&
  {Haehnelt}}{2006}]{VielHaehnelt2006}
{Viel} M.,  {Haehnelt} M.~G.,  2006, \mn@doi [\mnras]
  {10.1111/j.1365-2966.2005.09703.x}, \href
  {https://ui.adsabs.harvard.edu/abs/2006MNRAS.365..231V} {365, 231}

\bibitem[\protect\citeauthoryear{{Viel}, {Becker}, {Bolton}  \&
  {Haehnelt}}{{Viel} et~al.}{2013a}]{Viel2013}
{Viel} M.,  {Becker} G.~D.,  {Bolton} J.~S.,   {Haehnelt} M.~G.,  2013a,
  \mn@doi [\prd] {10.1103/PhysRevD.88.043502}, \href
  {https://ui.adsabs.harvard.edu/abs/2013PhRvD..88d3502V} {88, 043502}

\bibitem[\protect\citeauthoryear{{Viel}, {Schaye}  \& {Booth}}{{Viel}
  et~al.}{2013b}]{Viel2013_feedback}
{Viel} M.,  {Schaye} J.,   {Booth} C.~M.,  2013b, \mn@doi [\mnras]
  {10.1093/mnras/sts465}, \href
  {https://ui.adsabs.harvard.edu/abs/2013MNRAS.429.1734V} {429, 1734}

\bibitem[\protect\citeauthoryear{{Villasenor}, {Robertson}, {Madau}  \&
  {Schneider}}{{Villasenor} et~al.}{2022}]{Villasenor2022}
{Villasenor} B.,  {Robertson} B.,  {Madau} P.,   {Schneider} E.,  2022, arXiv
  e-prints, \href {https://ui.adsabs.harvard.edu/abs/2022arXiv220914220V} {p.
  arXiv:2209.14220}

\bibitem[\protect\citeauthoryear{{Walther}, {Hennawi}, {Hiss}, {O{\~n}orbe},
  {Lee}, {Rorai}  \& {O'Meara}}{{Walther} et~al.}{2018}]{Walther2018}
{Walther} M.,  {Hennawi} J.~F.,  {Hiss} H.,  {O{\~n}orbe} J.,  {Lee} K.-G.,
  {Rorai} A.,   {O'Meara} J.,  2018, \mn@doi [\apj] {10.3847/1538-4357/aa9c81},
  \href {https://ui.adsabs.harvard.edu/abs/2018ApJ...852...22W} {852, 22}

\bibitem[\protect\citeauthoryear{{Walther}, {O{\~n}orbe}, {Hennawi}  \&
  {Luki{\'c}}}{{Walther} et~al.}{2019}]{Walther2019}
{Walther} M.,  {O{\~n}orbe} J.,  {Hennawi} J.~F.,   {Luki{\'c}} Z.,  2019,
  \mn@doi [\apj] {10.3847/1538-4357/aafad1}, \href
  {https://ui.adsabs.harvard.edu/abs/2019ApJ...872...13W} {872, 13}

\bibitem[\protect\citeauthoryear{{Worseck}, {Prochaska}, {Hennawi}  \&
  {McQuinn}}{{Worseck} et~al.}{2016}]{Worseck2016}
{Worseck} G.,  {Prochaska} J.~X.,  {Hennawi} J.~F.,   {McQuinn} M.,  2016,
  \mn@doi [\apj] {10.3847/0004-637X/825/2/144}, \href
  {https://ui.adsabs.harvard.edu/abs/2016ApJ...825..144W} {825, 144}

\bibitem[\protect\citeauthoryear{{Wu}, {McQuinn}, {Kannan}, {D'Aloisio},
  {Bird}, {Marinacci}, {Dav{\'e}}  \& {Hernquist}}{{Wu} et~al.}{2019}]{Wu2019}
{Wu} X.,  {McQuinn} M.,  {Kannan} R.,  {D'Aloisio} A.,  {Bird} S.,  {Marinacci}
  F.,  {Dav{\'e}} R.,   {Hernquist} L.,  2019, \mn@doi [\mnras]
  {10.1093/mnras/stz2807}, \href
  {https://ui.adsabs.harvard.edu/abs/2019MNRAS.490.3177W} {490, 3177}

\bibitem[\protect\citeauthoryear{{Y{\`e}che}, {Palanque-Delabrouille}, {Baur}
  \& {du Mas des Bourboux}}{{Y{\`e}che} et~al.}{2017}]{Yeche2017}
{Y{\`e}che} C.,  {Palanque-Delabrouille} N.,  {Baur} J.,   {du Mas des
  Bourboux} H.,  2017, \mn@doi [\jcap] {10.1088/1475-7516/2017/06/047}, \href
  {https://ui.adsabs.harvard.edu/abs/2017JCAP...06..047Y} {2017, 047}

\bibitem[\protect\citeauthoryear{{{\v{D}}urov{\v{c}}{\'\i}kov{\'a}}, {Katz},
  {Bosman}, {Davies}, {Devriendt}  \&
  {Slyz}}{{{\v{D}}urov{\v{c}}{\'\i}kov{\'a}} et~al.}{2020}]{Durovcikova2020}
{{\v{D}}urov{\v{c}}{\'\i}kov{\'a}} D.,  {Katz} H.,  {Bosman} S. E.~I.,
  {Davies} F.~B.,  {Devriendt} J.,   {Slyz} A.,  2020, \mn@doi [\mnras]
  {10.1093/mnras/staa505}, \href
  {https://ui.adsabs.harvard.edu/abs/2020MNRAS.493.4256D} {493, 4256}

\makeatother
\end{thebibliography}

\bsp	
\label{lastpage}
\end{document}